\documentclass[prb,aps,showpacs,superscriptaddress,twocolumn,nofootinbib]{revtex4}

\usepackage{amsmath,amsfonts,amssymb}
\usepackage{graphicx}

\def \beq {\begin{equation}}
\def \eeq {\end{equation}}
\def \ba {\begin{eqnarray}}
\def \ea {\end{eqnarray}}
\newcommand{\upp}{\hspace{-0.2 pt}\uparrow}
\newcommand{\downn}{\hspace{-0.2 pt}\downarrow}

\newcommand{\ketbrad}[1]{|#1\rangle\!\langle #1|}
\newcommand{\ketbra}[2]{|#1\rangle\!\langle #2|}
\newcommand{\braket}[2]{\left\langle #1\vphantom{#2}\right.\left|\vphantom{#1}#2\right\rangle }

\newcommand{\mean}[1]{\langle#1\rangle}

\def\ket#1{\left| #1\right>}
\def\bra#1{\left< #1\right|}

\newcommand{\Epsilon}{\epsilon}

\newcommand{\SV}{\hat{\vec{S}}}
\newcommand{\Sz}{\hat{S}_z}

\newcommand{\Sx}{\hat{S}_x}

\newcommand{\Az}{\hat{A}_z}

\newcommand{\IV}{\hat{\vec{I}}}

\newcommand{\nv}{\vec{n}}

\newcommand{\Tc}{T_c}

\newcommand{\dBz}{\delta \hspace{-1pt} \hat{B}_z}

\def\3{2.8in}    
\def\2{2.5in}
\def\4{3.0in}

\begin{document}
\preprint{WHITEPAPER: not for distribution}

\title{Relaxation, dephasing, and quantum control of electron spins
  in double quantum dots}
\author{J. M. Taylor}
\affiliation{Department of Physics, Harvard
  University, 17 Oxford St., Cambridge, MA 02138 }\affiliation{Department of Physics,
  Massachusetts Institute of Technology, Cambridge, MA 02139}\author{J. R. Petta}\affiliation{Department of Physics, Harvard
  University, 17 Oxford St., Cambridge, MA 02138 }\affiliation{Department of Physics, Princeton University,
Princeton, New Jersey 08544}
\author{A. C. Johnson} \affiliation{Department of Physics, Harvard
  University, 17 Oxford St., Cambridge, MA 02138 } \author{A. Yacoby}\affiliation{Department of Physics, Harvard
  University, 17 Oxford St., Cambridge, MA 02138 }
\affiliation{Department of Condensed Matter Physics, Weizmann
  Institute of Science, Rehovot 76100, Israel}
\author{C. M. Marcus}\affiliation{Department of Physics, Harvard
  University, 17 Oxford St., Cambridge, MA 02138 }\author{M. D. Lukin}\affiliation{Department of Physics, Harvard
  University, 17 Oxford St., Cambridge, MA 02138 }
\begin{abstract}

  Recent experiments have demonstrated quantum manipulation of
  two-electron spin states in double quantum dots using electrically
  controlled exchange interactions.  Here, we present a detailed
  theory for electron spin dynamics in two-electron double dot systems
  that was used to guide those experiments and analyze the results.
  Specifically, we analyze both spin and charge relaxation and dephasing
  mechanisms that are relevant to experiments, and discuss practical
  approaches for quantum control of two-electron systems.  We show that
  both charge and spin dephasing play important roles in the dynamics
  of the two-spin system, but neither represents a fundamental limit
  for electrical control of spin degrees of freedom in semiconductor
  quantum bits.
\end{abstract}
\date{\today}
\maketitle



Electron spins in quantum dots represent a promising system for
studying mesoscopic physics, developing elements for
spintronics~\cite{awschalom01,awschalom07}, and creating building
blocks for quantum information
processing~\cite{loss98,imamoglu99,taylor05nature,hansonrmp06}. In
the field of quantum information, confined electron spins have
been suggested as a potential realization of a quantum bit, due to
their potential for long coherence
times~\cite{khaetskii00,khaetskii01,golovach04}. However, the
deleterious effects of hyperfine coupling to lattice nuclear
spins~\cite
{schulten78,burkard99,merkulov02,khaetskii02,erlingsson02,desousa03,coish04,yao05,deng05},
as found in experiments
\cite{bracker04,johnson05,koppens05,petta05,koppens06}, can
severely limit the phase coherence of electron spins.  Thus, it is
important to understand dynamics of electron spin coupled to
nuclei and to develop corresponding quantum control techniques to
mitigate this coupling.

Recent experiments by our group explored coherent spin
manipulation of electron spins to observe and suppress the
hyperfine interaction~\cite{johnson05,petta05,laird06}.  In this
paper, we present a detailed theory describing coherent properties
of coupled electrons in double quantum dots that was used to guide
those experiments and analyze the results.  The theory includes
hyperfine interactions, external magnetic field, exchange terms,
and charge interactions.

Our approach relies upon an approximation based on the separation
of time scales between electron spin dynamics and nuclear spin
dynamics. In particular, the time scales governing nuclear spin
evolution are slower than most relevant electron spin processes.
This allows us to treat the nuclear environment using a type of
adiabatic approximation, the quasi-static approximation
(QSA)~\cite{schulten78,merkulov02}.  In this model, the nuclear
configuration is fixed over electron spin precession times, but
changes randomly on the time scale over which data points in an
experiment might be averaged (current experiments acquire a single
data point on $\sim$ 100 ms timescales). We also consider the
first corrections to this approximation, where experimentally
relevant.

In what follows, we start by reviewing the theory of hyperfine
interactions in single and double quantum dots, focusing on
electrostatic control of electron spin-electron spin interactions.  We
then consider the role of charge dephasing and charge-based decay in
experiments involving so-called spin blockade, in which a simultaneous
spin flip and charge transition is required for electrons to tunnel
from one dot to another \cite{johnson05}.  Consistent with the
experiments, we find that blockade is reduced near zero magnetic field
over a range set by the average magnitude of the random Overhauser
(nuclear) field.  We then consider the effect of fast control of the
local electrostatic potentials of double quantum dots, and show how
this may be used to perform exchange
gates~\cite{loss98,burkard99,schliemann01}, and to prepare and measure
two-spin entangled states~\cite{taylor05,petta05}. Various limitations
to the preparation, manipulation, and measurement techniques, due to
nuclear spins, phonons, and classical noise sources, are considered.

Theories that explicitly include quantum mechanical state and
evolution effects of the nuclear spins both within and beyond the QSA
have been considered by several authors
(Refs.~\onlinecite{burkard99,khaetskii02,coish04,shenvi05,coish05,witzel05,yao05,deng05}).
Dephasing, decoherence, and gating error in double quantum dots have
also been investigated previously~\cite{coish05,hu05,klauser05}; the
present work develops the theory behind quantum control techniques
used in experiments, connecting the previous general theoretical
treatments to specific experimental observations.  The paper is
organized as follows.  Interactions of a single electron in a single
quantum dot, including hyperfine terms, are reviewed in
section~\ref{s:singledot}.  The quasi-static regime is defined and
investigated, and dephasing of electron spins by hyperfine
interactions in the quasi-static regime is detailed.  This provides a
basis for extending the results to double quantum dot systems.  We
then develop a theory describing the two-electron spin states of a double
quantum dot including the response of the
system to changes in external gate voltages, and the role of inelastic
charge transitions~\cite{brandes99,fujisawa01,brandes03}.  This is
combined with the theory of spin interactions in a single dot to
produce a theory describing the dynamics of the low energy states,
including spin terms, of the double dot system in two experimental
regimes.  One is near the charge transition between the two dots,
where the charge state of two electrons in one dot is nearly
degenerate with the state with one electron in each dot.  The other is
the far-detuned regime, where the two dots are balanced such that the
states with two electrons in either dot are much higher in energy.

In the remaining sections, we investigate situations related to the
experiments.  First we consider spin blockade near the charge
transition, as investigated in Ref.~\onlinecite{johnson05}, considering effects due to difference in dot sizes and expanding upon several earlier, informal ideas.  We then
analyze approaches to probing dephasing and exchange interactions, showing how errors effect fast gate control approaches for preparation and measurement
of two-electron spin states, as well as controlled exchange
interactions and probing of nuclear-spin-related dephasing, as
investigated in Ref.~\onlinecite{petta05}.  Finally, we consider
limitations to exchange gates and quantum memory of logical qubits
encoded in double dot systems~\cite{taylor05,taylor05nature}.

\section{Hyperfine interactions in a double quantum dot: a
  review \label{s:singledot}}

We begin by reviewing the basic physics of hyperfine interactions
for electron spins in single GaAs quantum dots. This section
reviews the established theory for single quantum
dots~\cite{schulten78,merkulov02,khaetskii02,desousa03b} and
considers dynamical corrections to the model of
Refs.~\onlinecite{schulten78,merkulov02}. This model will be used
in subsequent sections for the double-dot case.  Additional terms,
such as spin-orbit coupling, are neglected.
Theory~\cite{khaetskii00,khaetskii01,golovach04} and
experiment~\cite{hanson03,kroutvar04,johnson05} have demonstrated
that spin-orbit related terms lead to dephasing and relaxation on
time scales of milliseconds, whereas we will focus on interaction
times on the order of nanoseconds to microseconds.

\begin{figure}
\centering
\includegraphics[width=3.0in]{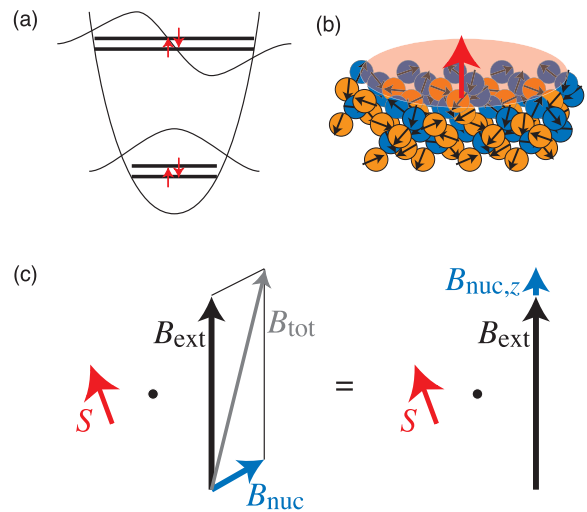}
\caption{ (a) A schematic potential and energy level diagram for a
single quantum dot in which one electron is confined to the low
energy spectrum of a three dimensional potential. Only the ground
and first-excited states, each a Kramer's doublet, are shown. (b)
The lowest orbital state has a spin-1/2 electron interacting with
the lattice nuclear spins. (c) Effective magnetic field due to
both external field and the nuclear field.  When the external
field is large, the transverse components of the nuclear field are
neglected in a rotating wave approximation. \label{f:singledot}}

\end{figure}

\subsection{Electron spin Hamiltonian for a single quantum dot}
The Hamiltonian for the Kramer's
doublet of the ground orbital state of the quantum dot
(denoted by the spin-1/2 vector $\SV$) including hyperfine contact
interactions with lattice nuclei (spins $\IV^{\beta,j}$) is~\cite{paget,merkulov02}
\beq H = \hbar \gamma_e \vec{B}_{\rm ext} \cdot \SV +  \hbar
\gamma_e \sum_{\beta,j} b_{\beta} \alpha_{j,\beta} \SV \cdot
\IV^{\beta,j} \label{e:ham} \eeq
where $\gamma_{e}=g^* \mu_B / \hbar$ is the gyromagnetic ratio for
electron spin $\SV$; sums are over nuclear species ($\beta$) and
unit cells ($j$). Correspondingly, $b_{\beta}$ is the effective
hyperfine field due to species $\beta$ within a unit cell, with
$b_{\rm ^{75}As} = -1.84$ T, $b_{\rm ^{69}Ga} = -1.52$ T, and
$b_{\rm ^{71}Ga} = -1.95$ T for GaAs~\cite{paget}. The coefficient
$\alpha_{j,\beta} = v_0 |\psi(\vec{r}_{j,\beta})|^2$ is the
probability of the electron being at unit cell $j$ (with nuclear
spin species $\beta$), $v_0$ is the volume of the unit cell (2
nuclei), and $\psi(r)$ is the envelope wavefunction of the
localized electron.

It is convenient to rewrite the Hamiltonian using a collective
operator for the nuclear spins,
$\hat{\vec{B}}_{\rm nuc} = \sum_{\beta} b_{\beta} \sum_j
\alpha_{j,\beta} \IV^{\beta,j}$.
This operator allows us to write the Hamiltonian as an electron spin
interacting with an external magnetic field, $\vec{B}_{\rm ext}$, and
an intrinsic field, $\hat{\vec{B}}_{\rm nuc}$:
\beq
H_{\rm
  eff} = \hbar \gamma_e (\vec{B}_{\rm ext} + \hat{\vec{B}}_{\rm
  nuc}) \cdot \SV~. \label{e:hambnuc}
\eeq

Several characteristic values~\cite{paget,merkulov02} for this
interaction  are noted in Table~\ref{t:scale}. The maximum nuclear
field value (all spins fully polarized with value $I=3/2$) is $h_0
= \sum_{\beta} b_\beta (x_\beta I^{\beta}) \sum_k \alpha_k$, where
we have separated out the relative population of nuclear species,
$x_{\rm ^{75}As} = 1$, $x_{\rm ^{69}Ga} = 0.6$, and $x_{\rm
^{71}Ga} = 0.4$ for GaAs, removing the $\beta$ dependence from the
$\alpha_{k,\beta}$.  This gives $b_0 = 5.3$ T. Second, when the
nuclear spins may be described by a density matrix $\rho =
\hat{1}/(2I+1)^N$ (infinite temperature approximation for $N$
nuclei), the root-mean-square (rms) strength of the
field\footnote{ A different convention than that of
  Ref.~\onlinecite{merkulov02} is used: gaussians are always described
  by their rms, rather than $\sqrt{2}$ times their rms.  Thus our
  value of $B_{\rm nuc} = \Delta_B/\sqrt{2}$, and similarly for other
  values to follow.  } is
\ba
B_{\rm nuc}& = & \sqrt{\mean{ |\hat{\vec{B}}_{\rm nuc}|^2}/3} =  \sqrt{ \sum_{\beta,k} b_{\beta}^2 \alpha_{k,\beta}^2 \mean{|\IV^{\beta,k}|^2}/3} \\
& = &\sqrt{(\sum_{\beta} x_\beta b_{\beta}^2) I(I+1) v_0/3 \int d^3r\ |\psi(r)|^4} \\
& = & h_1/\sqrt{N}\ .
\ea
where we have replaced $\sum_j v_0$ with $\int d^3r$. The
characteristic strength parameter is $h_1 = \sqrt{2I(I+1)/3
\sum_{\beta} x_\beta b_{\beta}^2} = 4.0$ T for GaAs, and $N$ is
defined as the number of nuclei with which the electron has
significant overlap, i.e., $N = 2 / [\int d^3r\ |\psi(r)|^4 v_0]$.
These numerical values are specific for GaAs quantum dots. Dots in
other materials with non-zero nuclear spin may be described by
similar parameters: a maximum field strength parameter $h_0$ and a
rms field strength parameter, $B_{\rm nuc} = h_1 / \sqrt{N}$.

\subsection{The quasi-static approximation for nuclear spins \label{s:singleDotDephasing}}

By writing the Hamiltonian (Eq.~\ref{e:hambnuc}) with nuclei as an
effective magnetic field, we have implicitly indicated that the field
may be considered on a similar footing to the external magnetic field.
In other words, the operator $\hat{\vec{B}}_{\rm nuc}$ may be replaced
by a random, classical vector $\vec{B}_{\rm nuc}$, and observables may
be calculated by averaging over the distribution of classical values.
The distribution in the large $N$ limit is
\beq
P(\vec{B}) =
\frac{1}{(2 \pi B_{\rm nuc}^2)^{3/2}} \exp(-(\vec{B} \cdot \vec{B})/2
B_{\rm nuc}^2)\ . \label{e:distro}
\eeq
This is the quasi-static
approximation (QSA) used in Refs.~\onlinecite{schulten78,merkulov02}\footnote{ The QSA is designated the
  quasi-stationary fluctuating fields approximation in
  Ref.~\onlinecite{merkulov02}.  }: we assume that over time scales
corresponding to electron spin evolution, the nuclear terms do not
vary.

In terms of dephasing, we cite the results of
Ref.~\onlinecite{merkulov02}.  At zero external magnetic field in
the Heisenberg picture, the electron spin $\SV$ evolves to: \beq
\mean{\SV(t)}_{\rm nuc} = \frac{\SV(0)}{ 3} (1 + 2(1 - (\gamma_e
B_{\rm nuc} t)^2) e^{-(\gamma_e B_{\rm nuc} t )^2}) . \eeq On the
other hand, at large external magnetic fields, $\Sz$ is conserved,
but transverse spin components (e.g., $\Sx$) decay as: \beq
\mean{\Sx}_{\rm nuc} = \frac{\Sx}{2}(1+e^{-\frac{1}{2}(\gamma_e
B_{\rm nuc} t)^{2}})\ . \eeq A time-ensemble-averaged dephasing
time due to nuclei in a single dot at large external magnetic
field (e.g., dot $i$) is
\beq
T_{2,i}^{*}=
\frac{1}{\gamma_e B_{{\rm nuc},i}}\ .
\eeq
This definition is appropriate when considering the decay of coherence
of a single electron in a single quantum dot.

Generalizing to all field values, for times longer than $T_2^*$,
\beq
 \SV(t\gg T_2^*) = \mean{(\SV(t=0) \cdot \nv) \nv}_{\rm nuc} \label{e:savg}
\eeq
is the average electron spin value, averaged over a time $\tau = 2
\pi / \omega$.

\begin{table*}

\begin{tabular}{|l|c|c|c|c|}
\hline Type & Time & Energy & Magnetic field & Typical value
\tabularnewline \hline \hline Charge & & & & \tabularnewline
\hline \ \ Charging energy &   & $E_c$ & & 5 meV   \tabularnewline
\ \ Orbital level spacing   &   & $\hbar \omega_0$ & & 1 meV
\tabularnewline \ \ Single dot two-electron exchange near $B_{\rm
ext} = 0$ &   & $\mathcal{J}$  & & 300 $\mu$eV \tabularnewline \ \
Double-dot tunnel coupling & & $\Tc$ & & 10 $\mu$eV
\tabularnewline \ \ Double-dot inelastic tunneling & &
$\Gamma(\epsilon)$ & & 0.01--100 neV \tabularnewline
\hline \hline 
Electron spin & & & & \tabularnewline
\hline
\ \ Larmor precession   & $t_L$ & $\hbar \gamma_e B_{\rm ext}$  & $B_{\rm ext}$  & 0--200 $\mu$eV \tabularnewline
\ \ Fully polarized overhauser shift & & $A I$ & $h_0$ & 130 $\mu$eV \tabularnewline
\ \ (Random) overhauser shift   & $T_2^*$   & $\hbar \gamma_e B_{\rm nuc}$   & $B_{\rm nuc} = \frac{h_1}{\sqrt{N}}$ & 0.1--1 $\mu$eV \tabularnewline
\hline \hline 
Nuclear spin species $\beta$  & & & & \tabularnewline
\hline
\ \ Larmor precession & $t_{nL,\beta}$ & $\hbar \gamma_{\beta} B_{\rm ext}$ & $B_{\rm ext}$  & 0--100 neV \tabularnewline
\ \ Knight shift & $t_{K,\beta}$ & $\hbar \gamma_e B_{\rm nuc}
\lambda_{\beta,j} \approx \frac{\hbar \gamma_e B_{\rm nuc}}{\sqrt{N}}$ & $\frac{\gamma_e B_{\rm nuc}}{\gamma_{\beta} \sqrt{N}}$ & 0.1--10 neV \tabularnewline
\ \ Dipole-dipole interaction (nearest neighbor) & $t_{dd}$ & $ \frac{(\hbar \gamma_{\beta})^2}{v_0}$ & $\frac{\hbar \gamma_{\beta}}{v_0}$ & 0.01 neV
\tabularnewline
\hline
\end{tabular}

\caption{\label{t:scale}
Time, energy, and magnetic field scales for electron and nuclear spins in a single and double quantum dots, from fast to slow.}

\end{table*}

At low magnetic fields, the QSA is valid up to the single electron
spin-nuclear spin interaction time ($O(\hbar N/A)$), which is of
order microseconds~\cite{merkulov02}.  In contrast, at large
external fields, the regime of validity for the QSA is extended.
Terms non-commuting with the Zeeman interaction may be eliminated
(secular approximation or rotating wave approximation), yielding
an effective Hamiltonian
\beq H_{\rm eff} = \hbar \gamma_e (B_{\rm ext} + B_{{\rm nuc}}^z)
\Sz\ . \label{e:heff} \eeq
The $z$ axis is set to be parallel to the external magnetic field.
Corrections to the QSA have a simple interpretation in the large
field limit.  As the Zeeman energy suppresses spin-flip processes,
we can create an effective Hamiltonian expanded in powers of
$1/B_{\rm ext}$ using a Schrieffer-Wolf transformation.  In the
interaction picture, we write the corrections to $H_{\rm eff}$ by
setting $B_{{\rm nuc}}^z = B_z + \dBz(t)$, where $B_z$ is the QSA
term, and $\dBz(t)$ are fluctuations beyond the QSA.  When the
number of nuclei $N$ is large and fluctuations small, we
approximate $\dBz$ by its fourier transformed correlation
function:
\beq
\mean{\dBz(t+\tau) \dBz(t)} = \int d\omega\ S(\omega) e^{i \omega \tau} \label{e:powerspec}
\eeq
where $S(\omega)$ has a high frequency cutoff $\gamma \ll \gamma_e
B_{\rm nuc}$.
The form of $S(\omega)$ depends on the detailed parameters of the
nuclear spin Hamiltonian and the nuclear spin-nuclear spin
interactions, and in general requires a many-body treatment.  A
variety of approaches have been used to successfully estimate
these
corrections~\cite{shenvi05,coish05,witzel05,yao05,deng05,witzel06}.
Any approach with an expansion in inverse powers of the external
field is compatible with our assumption of $S(\omega)$, provided
that the number of nuclear spins is sufficiently large that
Gaussian statistics may emerge.  In contrast, the validity of the
QSA in the low-field regime remains unproven, though recent
simulations~\cite{Al-Hassanieh06} suggest it may break-down before
$\hbar \sqrt{N}/A$ timescales.

\subsection{Hyperfine interactions in a double quantum dot}

We consider standard extensions to the single electron theoretical
model to describe the case of two electrons in adjacent, coupled
quantum dots, by considering only charge-related couplings, then
including spin couplings.  The relevant states are separated
electron states, in which one electron is in each quantum dot, and
doubly-occupied states, with two electrons in one of the two dots.

The doubly-occupied states are assumed to be singlets (appropriate
for small perpendicular magnetic field) \cite{burkard99}. The
higher excited states that are doubly-occupied are triplets with a
large energy gap $\mathcal{J}$.  This singlet--triplet energy gap
for doubly-occupied states facilitates elimination of the spin
interactions and the doubly-occupied triplet states. Furthermore,
by controlling the relative potential $\epsilon$ of the two
quantum dots using electrostatic potentials applied by external
gates, the ground state can be changed from one of the
doubly-occupied states to one of the separated electron states
(far-detuned regime on the other side of the charge transition)
\cite{vanderwiel}. Electrostatic control of the double dot
Hamiltonian will be analyzed in more detail in
sections~\ref{s:expt1}--\ref{s:echo}.

\begin{figure}
\centering
\includegraphics[width=\3]{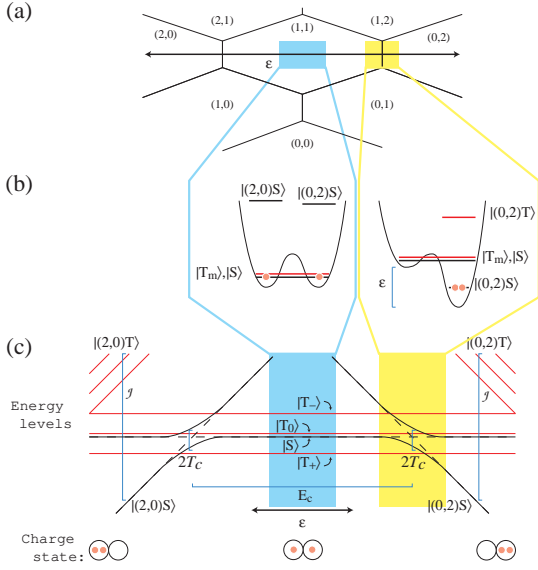}
\caption[Charge and orbital states for a double dot]{ (a) Charge
stability diagram for a double dot system. Double dot occupation
is denoted by ($n_l,n_r$). The detuning is parameterized by
$\epsilon$, and the far-detuned regime (light blue) and charge
transition (yellow) are shown. (b) Schematic of the double well
potentials along one axis ($x$) with tight confinement in the
other two axes (i.e., $y$ and $z$). In the far-detuned regime, the
(1,1) charge states are the ground state, while in the charge
transition regime, (0,2) can be the ground state. Triplet states
are indicated in red, while electron charges are indicated in
orange. (c) Energy level structure of the double dot system as a
function of detuning. From left to right the lowest energy charge
state as a function of $\epsilon$ is (2,0), (1,1), and (0,2). The
detuning at the middle of the graph corresponds to
$\epsilon=-E_c/2$, where $E_c$ is the charging energy of a single
dot. The three (1,1) triplet states (shown in red) are split by
Zeeman energy. \label{f:levels} }
\end{figure}

Formally, we eliminate all but one of the doubly-occupied states
following the prescription of Ref.~\onlinecite{coish05}. We
include the doubly-occupied state (0,2)S, where ($n_l,n_r$)
denotes number of electrons in left, right dots respectively, and
S denotes a singlet of electron spin, in addition to the singlet
and triplet manifolds of the (1,1) subspace. For notational
convenience, we set $\epsilon = 0$ to occur at the avoided
crossing between (1,1) and (0,2) in Fig.~\ref{f:levels}. There is
an avoided crossing at $\epsilon=0$ for the spin singlet manifold
due to quantum mechanical tunneling $\Tc$ between the two quantum
dots, while the spin triplet manifold is unaffected. The
Hamiltonian for the states $\ket{(0,2)S}$, $\ket{(1,1)S}$ can be
written as
\beq H_{11-02} = \left(\begin{array}{cc}
 -\epsilon & \Tc \\
 \Tc^* & 0 \\
 \end{array}\right)\ .
\label{e:h1102} \eeq
As the tunneling coefficient and external magnetic field are
assumed constant, we will choose $\Tc$ to be real by an
appropriate choice of gauge

For a slowly varying or time-independent Hamiltonian, the
eigenstates of Eq.~\ref{e:h1102} are given by
\ba
\ket{\tilde{S}} &= & \cos \theta \ket{S} + \sin \theta \ket{(0,2)S} \\
\ket{\tilde{G}} &= &-\sin \theta \ket{S} +\cos \theta \ket{(0,2)S}\ .
\ea
We introduce the tilde states as the adiabatic states, with
$\ket{\tilde{G}}$ the higher energy state.  The adiabatic angle is
$\theta = \arctan(\frac{2 \Tc}{\epsilon - \sqrt{4 |\Tc|^2+\epsilon^2}})$,
and the energies of the two states are
\ba
E_{\tilde{S}} & = & -\frac{\Tc}{2} \tan(\theta) \\ 
E_{\tilde{G}} & = & \frac{\Tc}{2} \tan(\pi/2-\theta) \ . 
\ea
When $\epsilon \ll -|\Tc|$, $\theta \rightarrow 0$,  the
eigenstates become $\ket{\tilde{S}}\rightarrow
\ket{S},\ket{\tilde{G}}\rightarrow \ket{(0,2)S}$.  For $\epsilon
\gg |\Tc|$, $\theta \rightarrow \pi/2$, and the eigenstates are
switched, with $\ket{\tilde{S}} \rightarrow \ket{(0,2)S}$ and
$\ket{\tilde{G}} \rightarrow \ket{S}$. As will be discussed later,
controllably changing $\epsilon$ allows for adiabatic passage
between the near degenerate spin states $\ket{S},\ket{T_m}$ (far
detuned regime) to past the charge transition, with $\ket{(0,2)S}$
as the ground state ($\epsilon \gg |\Tc|$). This adiabatic passage
can be used for singlet generation, singlet detection, and
implementation of exchange gates \cite{petta05}.

\begin{figure}
\centering
\includegraphics[width=\3]{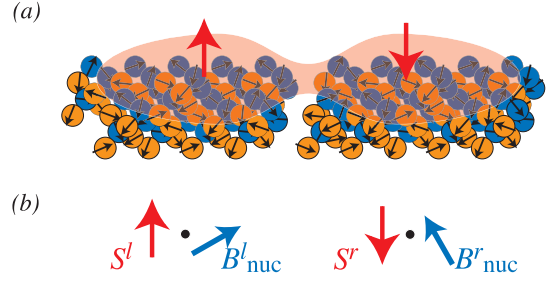}
\caption[A double quantum dot in the (1,1) configuration]{
A double quantum dot in the (1,1) configuration.  (a) Schematic of the
two-electron wavefunction in the far-detuned regime interacting with
lattice nuclear spins.  (b) Electron spins in the left and
right dots interacting with their respective effective nuclear fields  in the
quasi-static approximation.
\label{f:doubledot}}
\end{figure}

We now add spin couplings to the double dot system, including both
Zeeman interactions and hyperfine contact coupling.  Two effective
Hamiltonians, one for $\epsilon \ll -|\Tc|$ and one for $\epsilon
\sim 0$ are developed. Our approach is similar to that of
Ref.~\onlinecite{coish05}, and we include it here for
completeness. The spin interactions in a double quantum dot for
the states $\ket{T_m},\ket{\tilde{S}}$ may be written for
$\epsilon \ll -|\Tc|$
as
\beq H_{\rm hf,tot}  =  H_{\rm hf,eff}^l + H_{\rm hf,eff}^r -
J(\epsilon)\ketbrad{S} \label{e:hamhf} \eeq
where $l$ and $r$ refer to left and right dot, respectively, the
nuclear fields are determined by the ground orbital state envelope
wavefunctions of the single dot Hamiltonians (see Fig.~\ref{f:doubledot}), and $J(\epsilon) = - E_{\tilde{S}}(\epsilon)$.

Reordering terms
simplifies the expression:
\beq
H_{\rm hf, tot} = \hbar \gamma_e [\vec{B}\cdot(\SV^l + \SV^r) +
\vec{dB}\cdot(\SV^l - \SV^r)] - J(\epsilon)\ketbrad{S}\label{e:hftot}
\eeq
with an average field $\vec{B}=
\vec{B}_{\rm ext} + \frac{\vec{B}_{\rm
    nuc,l} + \vec{B}_{\rm nuc,r}}{2}$
and difference field $\vec{dB} =
(\vec{B}_{\rm nuc,l} - \vec{B}_{\rm nuc,r})/2$.  The form of
Eqn.~\ref{e:hftot} indicates that terms with $\vec{B}$ and $J(\epsilon)$ are
diagonal in total spin and spin projection along $\vec{B}$, creating a
natural set of singlet and triplet states. However, the term with
$\vec{dB}$ breaks total spin symmetry, and couples the singlet to the
triplet states.

\begin{figure}
\centering
\includegraphics[width=\3]{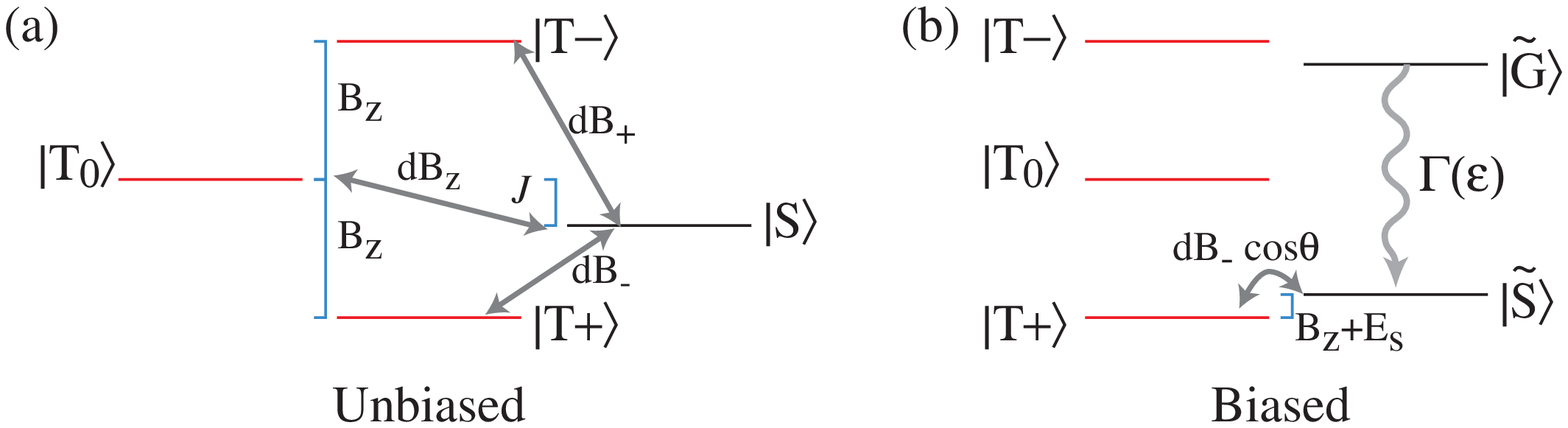}
\caption[Level structure: far-detuned and charge transition cases]{
(a) Levels in the far-detuned regime including all couplings of Eqn.~\ref{e:matrix}.
(b) Levels near the charge transition; the $\ket{T_+} \leftrightarrow
\ket{\tilde{S}}$ is near resonance, with the coupling between $\ket{T_+}$ and $\ket{\tilde{S}}$ indicated, as per Eqn.~\ref{e:ff}.
\label{f:spinLevels}}
\end{figure}

We can now write Eqn.~\ref{e:hftot} in matrix form in the basis
$\{\ket{T_+},\ket{T_0},\ket{T_-},\ket{S}\}$,
\beq
H = \hbar \gamma_e \left(
\begin{matrix}
B_z & 0 & 0 & \frac{dB_x - i dB_y}{\sqrt{2}} \\
0 & 0 & 0 & - dB_z \\
0 & 0 & -B_z & \frac{-dB_x - i dB_y}{\sqrt{2}} \\
\frac{dB_x + i dB_y}{\sqrt{2}} & - dB_z & \frac{-dB_x + i dB_y}{\sqrt{2}} & -J(\epsilon)/\gamma_e
\end{matrix}\right) \label{e:matrix}
\eeq
The corresponding level structure is given in
Fig.~\ref{f:spinLevels}a.  We have implicitly assumed the QSA in
writing this Hamiltonian by defining the axis of spin up and down as
$\vec{B}$, which is a sum of the external field and the average
nuclear field.  If the nuclear field fluctuates, those terms can
contribute by coupling different triplet states together.

With no external magnetic field, all states couple to the singlet,
and solving the dynamics requires diagonalizing the 4-by-4 matrix
of Eqn.~\ref{e:matrix}.  However, at finite magnetic field, a
large Zeeman splitting (which sets $B_z \gg B_{\rm nuc}$) allows
us to separate the system.  The far-detuned regime only has
transitions between the $m_s = 0$ states; in this basis
($\{\ket{T_0},\ket{S}\}$), the matrix becomes \beq H_{m_s=0} =
\hbar \gamma_e \left(
\begin{matrix}
0 & -dB_z \\
-dB_z & -J(\epsilon) /\gamma_e
\end{matrix} \right) \label{e:ms0} .
\eeq
This two-level system has appropriately straightforward dynamics, and
we investigate it in some detail below.

Near the charge transition and at finite magnetic field, another
coupling can occur, this time between $\ket{T_+}$ and
$\ket{\tilde{S}}$. This resonance corresponds to the adiabatic
singlet $\ket{\tilde{S}}$ having an exchange energy $J(\epsilon)$
close to the Zeeman-split triplet's Zeeman energy, $E_z = \hbar
\gamma_e B_z$.  We note that external magnetic field for GaAs will
be negative in this context (due to the negative electron
$g$-factor = -0.44). Written in the basis
$\{\ket{T_+},\ket{\tilde{S}}\}$, the Hamiltonian is \beq H_{\rm
flip-flop} = \hbar \gamma_e \left(
\begin{matrix}
B_z &  \frac{dB_x - i dB_y}{\sqrt{2}} \cos \theta \\
 \frac{dB_x + i dB_y}{\sqrt{2}} \cos \theta & -J(\epsilon) /\gamma_e
\end{matrix} \right) \label{e:ff} .
\eeq
We have subscripted Eqn.~\ref{e:ff} with ``flip-flop''  indicating that
flips between $\ket{\tilde{S}}$ and $\ket{T_+}$ result in the flipping
of a nuclear spin, which can be seen by identifying $dB_+ = dB_x + i dB_y
= (\hat{B}_{\rm nuc,l,+} - \hat{B}_{\rm nuc,r,+})/2$.

Because the $\tilde{S}$--$T_+$ resonance leads to spin flips and
eventual polarization of the nuclear field, the QSA will not be
valid if appreciable change of field occurs, and the overall
dynamics may go beyond the approximation. This has been examined
experimentally~\cite{ono03,koppens05} and
theoretically~\cite{taylor03b,erlingsson04,ramon06} for some
specific cases.  While the discussion to follow mentions this
resonance, it will focus on the zero field mixing of
Eqn.~\ref{e:hftot} and the far-detuned regime's finite field
mixing of Eqn.~\ref{e:ms0}.  We remark that Eqns.~\ref{e:ms0} and
\ref{e:ff} have been previously derived outside of the
QSA~\cite{coish05}.

We have now established that in the far-detuned regime, the
relevant spin interactions are limited to dynamics within the
singlet-triplet subspace and determined by the Hamiltonian in
Eqn.~\ref{e:ms0}. Similarly, near the charge transition a
resonance between $\ket{S}$ and $\ket{T_+}$ may be observed; as
this resonance allows for nuclear spin polarization, it may only
be partially described by the QSA, and we do not consider its
dynamics in detail.  However, we note that in the absence of
nuclear spin polarization the resonance occurs when the Zeeman
splitting of the external field equals the exchange energy,
$J(\epsilon)$.  Thus, if the Zeeman energy is known, measuring the
position of the splitting gives a map between external parameters
and the actual exchange energy.

\section{Nuclear-spin-mediated relaxation in double dots \label{s:expt1}}

In this section we consider the case in which the ground state of
the system is $\ket{(0,2)S}$ and the low-lying excited states are
the (1,1) states, $\{ \ket{S}, \ket{T_m}\}$.  This situation
occurs in dc transport when the system is in the spin blockade
regime, where transitions from $\ket{T_m}$ to $\ket{(0,2)S}$ are
suppressed because they require both a spin and charge transition.
Previous theoretical work for two electron systems has focused on
triplet and singlet decay of two-electron states in a single
quantum dot~\cite{erlingsson01}; in contrast, the present analysis
deals with a double quantum dot system where the electrons can be
well separated. Contrary to more general spin blockade
calculations\cite{jouravlev05} and experiments, the present work
is focused entirely on the rate limiting step of blockade: the
spin flip followed by charge transition within the double quantum
dot. Several groups\cite{hanson03,elzerman04,kroutvar04} have
studied spin relaxation between Zeeman split spin states at high
magnetic field ($B>$4 T). The measured relaxation rates were found
to scale as $B^4$, consistent with a spin-orbit mediated spin
relaxation process\cite{golovach04}. Similarly, single dot
measurements of triplet-singlet relaxation when $\mathcal{J} \gg
\gamma_e B_{\rm nuc}$ (i.e., when the effect of nuclei is small)
indicate long lifetimes, likely limited by similar spin-orbit
mediated mechanisms or cotunneling to the
leads\cite{fujisawa01,meunier06}. On the other hand, at low field
and small exchange, when the splitting between spin states becomes
comparable to $B_{\rm nuc}$, the hyperfine interaction
dramatically increases the spin relaxation rate. Recent
experiments have measured spin relaxation between nearly
degenerate singlet and triplet spin states in this
regime~\cite{johnson05,hanson05}.  Experimental techniques are
discussed in Ref.  \cite{petta05b}, and a full analysis of the
field and energy dependence of the relaxation rate is discussed in
Ref. \cite{johnson05}. We only briefly outline the salient
features of experiment, focusing instead on developing a more
rigorous basis for the theory of previous, published work.

\begin{figure}
\centering
\includegraphics[width=\3]{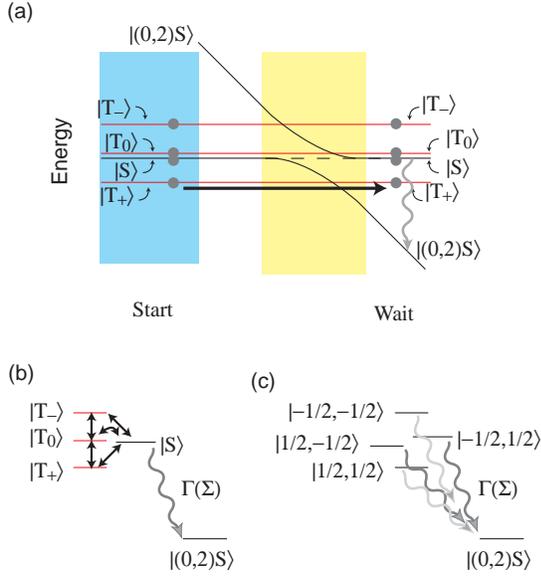}
\caption[Energy level structure as a function of detuning]{ (a) Energy level structure as a function of detuning.
  Coupling to a Fermi sea with $k_b T \gg \gamma_e |B_{\rm ext}|$
  leads to equal filling of all four low energy states in the far-detuned
  regime (labelled start).  Then $\epsilon$ is changed rapidly with
  respect to the tunnel coupling, leading to all four spin states
  still in the (1,1) charge configuration.  The time spent waiting in
  this configuration results in slow decay of the metastable (1,1)
  states to the $\ket{(0,2)S}$ state.  (b) Nuclear spins couple
  between the eigenstates of exchange, and slow inelastic decay at a
  rate $\Gamma$ proceeds from $\ket{S}$ to $\ket{(0,2)S}$.  (c) The
  same process, but for the eigenstates of the double dot Hamiltonian.
  The $m_s=0$ states are equal superpositions of $\ket{S}$ and decay
  rapidly to $\ket{(0,2)S}$, while the $|m_s|=1$ states are only
  weakly mixed at large magnetic field with the $\ket{S}$, resulting
  in slow decay to $\ket{(0,2)S}$.
\label{f:alexExpt}}
\end{figure}

Experiments are performed near the two-electron regime with very
weak tunnel coupling so that $\Tc$ is slower than the pulse rise
times ($\Tc \ll 1 \mu$eV). Pulsed-gate techniques are used to
change the charge occupancy from (0,1) to (1,1) to (0,2) and back
to (0,1). In the (1,1) charge configuration with weak interdot
tunnel coupling the $\ket{S}$ and $\ket{T_m}$ states are nearly
degenerate. Shifting the gates from (0,1) to (1,1) creates a
mixture of all four states, $\ket{S},\ket{T_{m=-1,0,1}}$ by
loading an electron from a nearby Fermi sea.  Then, the system is
rapidly (non-adiabatic with respect to tunnel coupling $\Tc$)
shifted to the (0,2) regime, with $\ket{(0,2)S}$ as the ground
state.  In this rapid shift procedure, the singlet $\ket{S}$ does
not adiabatically follow to the doubly-occupied singlet
$\ket{(0,2)S}$, but instead follows the Zener branch of the
avoided crossing and stays in $\ket{S}$, as is illustrated in
Fig.~\ref{f:alexExpt}.

Past the charge transition, when the adiabatic basis
$\ket{\tilde{S}},\ket{\tilde{G}}$ is an appropriate representation
of the system, it is possible for the system to experience
inelastic decay from the excited state $\ket{\tilde{G}}$ to the
ground state $\ket{\tilde{S}}$ via charge coupling, e.g., to
phonons.  The energy gap for $\epsilon \gg |\Tc|$ is
$E_{\tilde{G}}-E_{\tilde{S}} = \sqrt{\epsilon^2 + 4
  |\Tc|^2} \approx \epsilon$.
Inelastic decay near a charge transition in a double
quantum dot has been investigated in great
detail~\cite{fujisawa98,brandes99,brandes03}, and we do not seek to
reproduce those results.  Instead, we note that the decay from the
excited state $\ket{\tilde{G}}$ to the ground state $\ket{\tilde{S}}$
is well described by a smoothly varying, energy-dependent decay rate
$\Gamma_{ph}(\Epsilon)$.  Incoherent population of $\ket{\tilde{G}}$
by absorption of a thermal phonon is suppressed as long as $2 \Tc >
k_b T$, which is satisfied for $\Tc = 0.01$ meV and $T = 100$ mK.

Finally, we can combine the coherent spin precession due to interaction with nuclear spins  with the charge-based decay and dephasing mechanisms to
investigate relaxation of $\ket{T_m}$ states to the state
$\ket{\tilde{S}}$.  Of particular interest is the regime past the
charge transition,
$\epsilon \gg \Tc$, where $\ket{\tilde{G}} \approx \ket{S}$ becomes
nearly degenerate with $\ket{T_0}$, as was studied in the experiment
of Ref.~\onlinecite{johnson05}.  An effective five-level system is formed
with the levels $\{ \ket{T_+},\ket{T_0},\ket{T_-},\ket{S} \}$
described by the spin Hamiltonian of Eqn.~\ref{e:matrix}, while
inelastic decay from $\ket{S}$ to $\ket{(0,2)S}$ (the fifth level) is
possible at a rate $\Gamma(\Epsilon)$, as shown in Fig.~\ref{f:alexExpt}b.

To analyze this process, we start with the Louivillian superoperator
that describes inelastic tunneling:
\ba
\dot \rho & = & i [\rho,H/\hbar] + \Gamma(\Epsilon)/2 \Big[ \ketbrad{S}\rho + \rho \ketbrad{S}
 \nonumber \\
 & & \ \
- 2 \ketbra{(0,2)S}{S} \rho \ketbra{S}{(0,2)S} \Big] \label{e:L}
\ea
where
\beq
H  = \hbar  \gamma_e [ \vec{B}^l \cdot \SV^l + \vec{B}^r \cdot \SV^r ]_{(1,1)} - \hbar \Epsilon \ketbrad{(0,2)S} \ .
\eeq
$l$ and $r$ indicate left and right spins for the (1,1) charge
space. Assuming the nuclear field is quasi-static (QSA), we can diagonalize $H$.  The eigenstates are the
ground state, $\ket{(0,2)S}$, and (1,1) states with spin aligned and
anti-aligned with the local magnetic fields, $\vec{B}^{l,r} =
\vec{B}_{\rm ext} + \vec{B}_{\rm nuc}^{l,r}$.  We write these
eigenstates as $\ket{s,s'} = \ket{s}_l \otimes \ket{s}_r$, where
$s,s'=\pm 1/2$ are the eigenvalues of the spin projection on the
fields of the $l$ and $r$ dots, respectively.  The eigenvalue for
$\ket{(0,2)S}$ is  $E_G = -\Epsilon$ and the other four eigenstates
$\ket{s,s'}$ have energy
\beq
E_{s,s'} = s \gamma_e |B^l| +
s' \gamma_e |B^r|\ . \label{e:Esubs}
\eeq

In considering decay from the energy eigenstates of the nuclear field,
$\ket{s,s'}$ to $\ket{(0,2)S}$, we eliminate rapidly
varying phase terms, e.g., $\ket{1/2,-1/2} \bra{-1/2,1/2}$.  This is appropriate provided that
the inelastic decay mechanism, $\Gamma(\Epsilon)$, is
slow in comparison to the electrons' Larmor precession in the nuclear
field $B_{\rm nuc}$.  In this limit, each state $\ket{s,s'}$ decays to
$\ket{(0,2)S}$ with a rate given by $\Gamma(\Epsilon)
|\braket{s,s'}{(0,2)S}|^2$, as indicated in Fig.~\ref{f:alexExpt}c.  A
detailed analysis is given in appendix~\ref{s:inelastic}.  For
convenience, we write $c_{s,s'} = \braket{s,s'}{(0,2)S}$.

Starting with a mixed state of the (1,1) subspace (as in
Ref.~\onlinecite{johnson05}), we can find analytical expressions
for the time evolution of the density matrix of an initial form
$\rho(t=0) = \sum_{s,s'} \ketbrad{s,s'}/4$.  This initial state
corresponds to a mixture of the four (1,1) spin states.  The
charge measurement distinguishes only between (1,1) states and
$\ket{(0,2)S}$; accordingly, we evaluate the evolution of the
projector for the (1,1) subspace $P_{11} = \sum_{s,s'}
\ketbrad{s,s'}$.  In particular, \beq P_{11}(t) =
e^{-\Gamma(\Epsilon) |c_{++}|^2 t}/2 + e^{-\Gamma(\Epsilon)
|c_{+-}|^2 t}/2 \eeq.

Finally, we must average over
possible initial nuclear spin configurations to find the measured
signal. This means evaluating $\mean{P_{11}(t)}_{\rm nuc}$, a
difficult task in general. However,
\beq
\mean{P_{11}(t)}_{\rm nuc} \approx e^{-\Gamma(\Epsilon) \mean{|c_{++}|^2}_{\rm nuc} t}/2 + e^{-\Gamma(\Epsilon)  \mean{|c_{+-}|^2}_{\rm nuc} t}/2\ .
\eeq
In this approximation, we replace the average of the exponents with
the average values for the coefficients $|c_{++}|^2, |c_{+-}|^2$.  The
validity of this approximation can be checked with numerical
integration, and for the range of parameters presented here the
approximation holds to better than 1\%.

\begin{figure}
\centering
\includegraphics[width=\3]{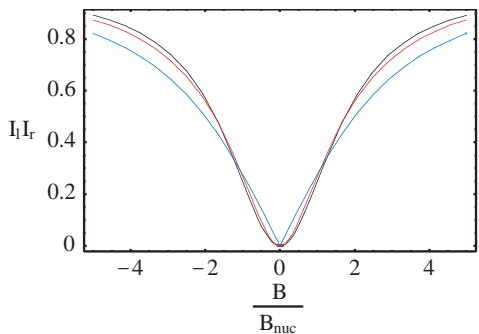}
\caption[Field dependence of spin-blockade]{ The product $I_l I_r$ as a function of external magnetic
  field $B$ in units of average nuclear field, $B_{\rm nuc} = (B_{{\rm
      nuc},l} + B_{{\rm nuc},l})/2$.  Several ratios of dot nuclear
  fields, $r=B_{{\rm nuc},l}/B_{{\rm nuc},r}$ are considered: $r=1$
  (black), $r=1/3$ (red) and $r=0$ (blue). Cusp behavior near zero
  field is found in the limit of highly inhomogeneous dot field
  strengths.  }
\label{f:overlapwithfield}
\end{figure}

The mean values of the coefficients $|c_{++}|^2, |c_{+-}|^2$ are in
turn straightforward to calculate approximately (as done in
Ref.~\onlinecite{johnson05}, supplemental information\footnote{ In
  contrast to Ref.~\onlinecite{johnson05}, where $B_{\rm nuc} =
  \sqrt{\mean{|\hat{\vec{B}}_{\rm nuc}|^2}}$, we define $B_{\rm nuc} =
  \sqrt{\mean{|\hat{\vec{B}}_{\rm nuc}|^2}/3}$.
} ), giving
\ba
\Gamma_{+-} = \Gamma(\Epsilon) \mean{|c_{+-}|^2}_{\rm nuc} & = & \frac{\Gamma(\Epsilon)}{4} (1+ I_l I_r) \label{e:Gamma1}\\
\Gamma_{++} = \Gamma(\Epsilon) \mean{|c_{++}|^2}_{\rm nuc} & = &
\frac{\Gamma(\Epsilon)}{4} (1 - I_l I_r) \label{e:Gamma2}
\ea
with
\beq
I_{l} = \frac{1}{\sqrt{[1+3 (\frac{B_{{\rm nuc},l}}{B_{\rm ext}})^2]}}
\eeq
and similarly for $I_r$.  We plot the product $I_l I_r$, found in both
Eqns.~\ref{e:Gamma1}-\ref{e:Gamma2}, as a function of external
magnetic field for increasing difference in dot sizes in
Fig.~\ref{f:overlapwithfield}.  This indicates that the effective
decay rates are largely independent of the ratio of dot sizes, relying
only on the average effective nuclear field, $B_{\rm nuc} = (B_{{\rm
    nuc},l} + B_{{\rm nuc},r})/2$.

We find that past the charge transition with $\epsilon \gg |\Tc|$
the states $\ket{S}$ and $\ket{T_0}$ decay to $\ket{(0,2)S}$ with
a lifetime $\Gamma_{+-}^{-1}$, while the two $|m_s|=1$ triplet
states have a lifetime $\Gamma_{++}^{-1}$.  At finite magnetic
field, $\Gamma_{+-} \gg \Gamma_{++}$, and we can call the states
of the $|m_s|=1$ subspace ``metastable''.  The metastability
allows for charge-based measurement to distinguish between
$\{\ket{S},\ket{T_0}\}$ and $\ket{T_{\pm}}$ subspaces: by using a
nearby charge sensor, the decay of $\ket{S},\ket{T_0}$ may be
detected long before $\ket{T_{\pm}}$ has finite probability of
decay in the weak tunneling limit.  This indicates that, while at
zero field decay of the (1,1) states to the state $\ket{(0,2)S}$
is governed by a single exponential, a double-exponential behavior
appears as $B > B_{\rm nuc}$ is satisfied, in direct confirmation
of the results of Ref.~\onlinecite{johnson05}.

Contrary to expectation, the blockade is contributed to solely by
the $|m_s|=1$ triplet states. In particular, spin blockade is
charge transport at finite bias through, for example, the charge
states $(0,1) \rightarrow (1,1) \rightarrow (0,2)$. For biases
between left and right leads that are less than $\mathcal{J}$ only
the four (1,1) spin states and the state $\ket{(0,2)S}$ are
necessary for understanding the process.  An electron (of
arbitrary spin) loads from the left lead, creating with equal
probability any of the states $\ket{s,s'}$.  This then tunnels
with a rate $\Gamma_{+-}$ or $\Gamma_{++}$ to the state
$\ket{(0,2)S}$, after which the extra electron on the right
tunnels into the leads, and the cycle repeats anew.  The average
current through the device is dominated by the slowest rate, which
in the absence of cotunneling, is $\Gamma_{++}$.  In other words,
loading into a spin-aligned state $\ket{s,s}$ prevents further
charge transport until it decays, with rate $\Gamma_{++}$, or is
replaced from the leads by a cotunneling process.

The measurements by Johnson \textit{et al.} demonstrate that the
transition probability from (1,1) to $\ket{(0,2)S}$ depends
strongly on both magnetic field and detuning \cite{johnson05}. Our
theoretical model, which accounts for hyperfine mixing coupled
with inelastic decay agrees well with experimental results for
timescales less than 1 ms. The discrepancy between experiment and
theory for longer times suggests that other spin relaxation
processes may become important above 1 ms (spin-orbit).

\section{Quantum control of two electron spin states
\label{s:expt2}}

We now analyze how time-dependent control of gate parameters
(e.g., $\epsilon$) may be used to control electron spin in double
quantum dots. Of particular interest are methods for probing the
hyperfine interaction more directly than in the previous section.
The new techniques we use are primarily rapid adiabatic passage
and slow adiabatic passage.  Rapid adiabatic passage (RAP) can
prepare a separated, two-spin entangled state ($\ket{S}$ in the
far-detuned regime), and when reversed, allows a projective
measurement that distinguishes the state $\ket{S}$ from the
triplet states, $\ket{T_m}$.  A similar technique used at large
external magnetic field, slow adiabatic passage (SAP), instead
prepares and measures eigenstates of the nuclear field,
$\ket{s,-s}$. We connect these techniques with the experiments in
Ref.~\onlinecite{petta05} and estimate their performance.

\subsection{Spin-to-charge conversion for preparation and measurement}

Adiabatic passage from $\epsilon \ll -\Tc$ to $\epsilon \gg \Tc$
maps the far-detuned regime states $\ket{S},\ket{T_m}$ to the
states past the charge transition $\ket{(0,2)S},\ket{T_m}$,
allowing for a charge measurement to distinguish between these
results~\cite{taylor05,petta05}.  In the quantum optics
literature, when adiabatic transfer of states is fast with respect
to the relevant dephasing (nuclear spin-induced mixing, in our
case), it is called ``rapid adiabatic passage'' (RAP) and we adopt
that terminology here.

\begin{figure}
\centering
\includegraphics[width=\3]{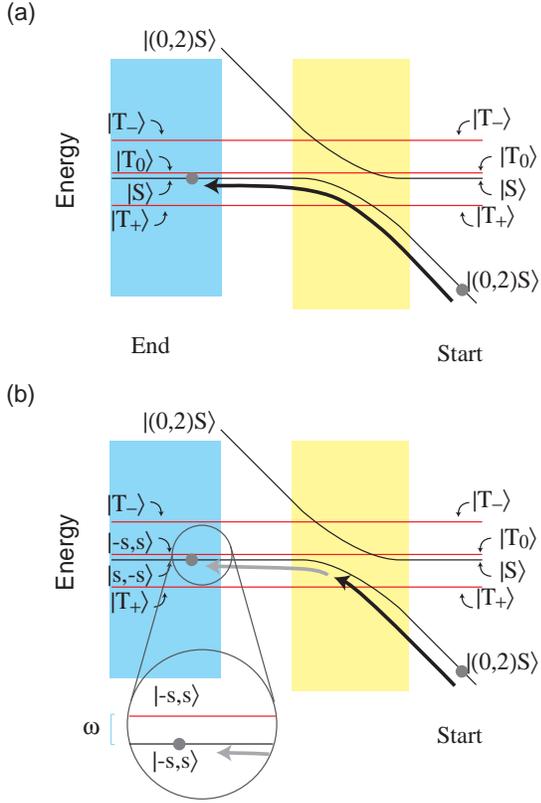}
\caption[Rapid- and slow-adiabatic passage]{ (a) Rapid-adiabatic
passage: starting in the state $\ket{(0,2)S}$, the detuning is
changed from $\epsilon \gg \Tc$ to $\epsilon \ll - \Tc$, fast with
respect to the nuclear energy scale, $\gamma_e B_{\rm nuc}$. (b)
Slow adiabatic passage: as above, but once the system is past the
$S-T_+$ degeneracy point, the change of $\epsilon$ is made slow
with respect to the nuclear energy scale. The zoomed-in section
shows the current nuclear energy splitting ($\hat{\omega}$) and
the nuclear field eigenstates, $\ket{s,-s}$ and $\ket{-s,s}$.
Both procedures may be reversed to transfer either $\ket{S}$ (RAP)
or $\ket{s,-s}$ (SAP) to $\ket{(0,2)S}$ while keeping the other
states within the (1,1) charge configuration, allowing for
charge-based measurement of the system. \label{f:RAPSAP} }
\end{figure}

When the change of detuning, $\epsilon$, is adiabatic with respect
to tunnel coupling, $\Tc$, but much faster than $\gamma_e B_{\rm
nuc}$ (the hyperfine coupling), the adiabatic passage is
independent of the nuclear dynamics. For example, starting past
the charge transition with the state $\ket{(0,2)S} (=
\ket{\tilde{S}})$ and $\epsilon \gg \Tc$ and using RAP to the
far-detuned regime causes adiabatic following to the state
$\ket{S}$.  This prepares a separated, entangled spin state.  The
procedure is shown in Fig.~\ref{f:RAPSAP}a.

The reverse procedure may be used to convert the singlet state to
the charge state (0,2) while the triplet remains in (1,1).  Then,
charge measurement distinguishes singlet versus triplet.
Specifically, if we start with some superposition in the
far-detuned regime, $\ket{\psi} = c_S \ket{S} + \sum_m c_m
\ket{T_m}$ where $c_{S},c_{m}$ are quantum amplitudes, after RAP
past the charge transition, with $\epsilon \gg \Tc$, the state is
$\ket{\psi'} = c_S e^{i \phi} \ket{(0,2)S} + \sum_m c_m
\ket{T_m}$, where $\phi$ is the adiabatic phase accumulated.
Recalling that $\ket{(0,2)S}$ is in the (0,2) charge subspace,
while $\ket{T_m}$ states are in the (1,1) subspace, a nearby
electrometer may distinguish between these two results, performing
a projective measurement that leaves the $\ket{T_m}$ subspace
untouched. Furthermore this measurement is independent of the
adiabatic phase.

\begin{figure}
\centering
\includegraphics[width=\2]{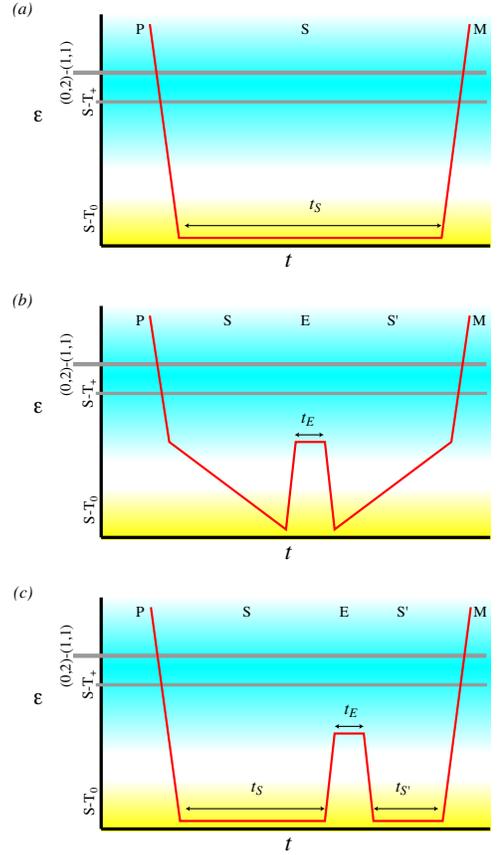}
\caption[Pulse sequences]{Pulse sequences: detuning parameter $
\epsilon$ versus time, for (a)
   RAP for measurement of the singlet autocorrelation function, (b) SAP
   for the
   exchange-gate sequence, and (c) the singlet-triplet echo
sequence.  Blue is the charge transition region, while yellow is
the far-detuned regime.  The charge degeneracy point (the crossing
from (1,1) to (0,2)) and the degeneracy between $\ket{\tilde{S}}$
and $\ket{T_+}$ (when $j(\epsilon) = \gamma_e |B_{\rm ext}|$) are
shown for reference. \label{f:pulses}}
\end{figure}

In contrast, if the change of $\epsilon$ is slow with respect to
nuclei, adiabatic passage follows to eigenstates of the hyperfine
interaction. For simplicity, we assume that RAP is used between
the charge transition to just past the $S-T_+$ resonance of
Eqn.~\ref{e:ff}, such that we may neglect transfer between the $S$
and $T_+$ states (see also Fig.~\ref{f:pulses}b.  This requires an
external magnetic field $|B_{\rm ext}| \gg B_{\rm nuc}$.
Continuing from this ($\epsilon < -\Tc$) point to the far-detuned
regime, $\epsilon$ is changed {\rm slowly} with respect to $\hbar
\gamma_e B_{\rm nuc} $.  Accordingly, adiabatic passage proceeds
into eigenstates of the {\em nuclear} field, $\ket{s,-s}$, as
shown in Fig.~\ref{f:RAPSAP}b.   These are the product spin
states, with one spin up and the other down with respect to
magnetic field. This technique may be called slow adiabatic
passage (SAP).

Rapid adiabatic passage maps $\ket{S} \leftrightarrow e^{i \phi}
\ket{(0,2)S}, \ket{T_m} \leftrightarrow \ket{T_m}$, leaving the
triplet states unperturbed. For SAP, the mapping is
\beq \left \{ \begin{array}{ccc}
     \ket{(0,2)S} & \leftrightarrow & e^{i \phi} \ket{s,-s} \\
     \ket{T_0} & \leftrightarrow & e^{i \phi'} \ket{-s,s} \\
     \ket{T_{\pm 1}} & \leftrightarrow & \ket{T_{\pm 1}}
\end{array} \right.
\eeq
It is always the current, lowest energy eigenstate of the nuclear
field that $\ket{(0,2)S}$ maps to.  That is, we choose $s$ such
that $E_{s,-s} < E_{-s,s}$, with $E_{s,-s} = s \gamma_e (B^l_{{\rm
nuc},z} - B^r_{{\rm nuc},z})$, (see Section~\ref{s:expt1},
Eqn.~\ref{e:Esubs}, evaluated at large external magnetic field).
We remark that SAP allows for deterministic preparation and
measurement of states rotated $\pi/2$ with respect to
$\ket{S},\ket{T_0}$ without direct knowledge of which states they
correspond to for each realization.

We now examine the adiabaticity condition for slow adiabatic
passage. In SAP, $\epsilon$ is changed at a constant rate to
approach the far-detuned regime (point S). Using the approximate
relation $J(\epsilon)=4\Tc^{2}/\epsilon$, the adiabaticity
condition is $\hbar\dot{J}\ll(\hbar\omega)^{2}$. Neglecting
factors of order unity, this can be rewritten
$\hbar\dot{\epsilon}\frac{\Tc^{2}}{\epsilon^{2}}\ll(\hbar\omega)^{2}$.

As a specific case, we consider $\Tc\simeq5\ \mu$eV, and $\epsilon
\in[50,550]\ \mu$eV.  The required time to make the 500 $\mu$eV
change, $\tau$, gives $\hbar\dot{\epsilon}=\frac{\hbar}{\tau}500\
\mu$eV, and roughly,
$\hbar\dot{\epsilon}\frac{\Tc^{2}}{\epsilon^{2}}=\frac{\hbar}{\tau}
\times50$ neV. In units of time, $\hbar^{2}(\tau\times13\ {\rm
ns})$. For $\tau=1000$ ns, the adiabaticity requirement is that
the current value of
$1/|\omega|\lesssim3\times\sqrt{1000\times13}\ {\rm ns}=\ 300$ ns.
For the nuclear fields in lateral quantum dots such as those of
Ref.~\onlinecite{johnson05} (each dot with a $T_{2}^{*}=10$ ns),
the probability of $1/|\omega|>300$ ns is
\beq P(|\omega|<3\ \mu{\rm s}^{-1})  = 2\int_{0}^{3\ \mu{\rm
s}^{-1}}d\nu\ \frac{e^{-(\nu
T_{2}^{*})^{2}}}{\sqrt{\pi(T_{2}^{*})^{-2}}} \eeq
This gives an error probability of 3\% for 300 ns rise time, that
is, every 1 in 30 experimental runs, the nuclear gradient will be
too small for the adiabatic filling of $\ket{\pm}$ to occur.

We can ask the effect of finite, residual exchange energy $J_{{\rm
res}}$ at the S point. Finite $J$ leads to filling of a
superposition of $ \ket{s,-s}$ and $\ket{-s,s}$: \beq
\cos(\phi)\ket{s,-s}-\sin(\phi)\ket{-s,s}~, \eeq
where the value of $s$ is, as above, determined by the current
value of $\omega$ and $\phi=\arctan[J_{{\rm res}}/(\sqrt{J_{{\rm
       res}}^{2}+\omega^{2}}+|\omega|)]$ is the adiabatic angle. The
resulting loss of contrast will be
$\sin^{2}(\phi)\simeq(J/2\omega)^{2}$. For residual $J_{{\rm
     res}}\sim0.1\ \Omega$, the error is less than 1\%. For $J_{{\rm
     res}}\sim0.25\ \Omega$ the error is order 2\%.
The role of residual nuclear fields during the exchange gate is
evaluated elsewhere~\cite{taylor05spinbath}.

\subsection{Probing the nuclear field and exchange interactions with
adiabatic passage}

We now consider how adiabatic passage can be used to probe the
dephasing and exchange energy of a spin-singlet state.  This
relates directly to a critical question in quantum information
science: how long can two electron spins remain entangled when the
electrons are spatially separated on a GaAs chip. In our model,
variations in the local nuclear environment cause the spatially
separated electrons to experience distinct local magnetic fields,
and hence precess at different rates, mixing the singlet and
triplet states.  If many measurements are taken to determine the
probability of remaining a singlet, the time-ensemble-averaging
leads to an observable dephasing of the singlet state
($T_2^*$)~\cite{petta05}.

To evaluate the effects of nuclei on this process, we will
calculate the singlet autocorrelation function $A_{S}(t) =
|\braket{S(t)}{S(0)}| ^2$ for the far-detuned regime.  This
autocorrelation function has been evaluated for quantum
chemistry~\cite{schulten78}, but not for this specific scenario.
We remark that our approach is similar to the single dot case
considered in Refs.~\onlinecite{merkulov02,khaetskii02}.

We start by evaluating the evolution operation $U(t)$, where the
Schrodinger picture $\ket{S(t)} = U(t) \ket{S}$. Taking
$J\rightarrow 0$ for the far-detuned regime, we solve analytically
the equation of motion any spin state of the (1,1) subspace. In
particular, we take the Hamiltonian of Eqn.~\ref{e:hamhf} and
write it in the form two effective fields, each acting separately
on one spin. The evolution operator, $U(t)=\exp(-iHt/\hbar)$ can
be factorized as $U(t)=U_{l}(t)\otimes U_{r}(t)$, where
\beq U_{i}(t)=\exp(-i\gamma_e t[\vec{B}+\vec{B}_{{\rm
nuc},i}]\cdot\vec{S}_ {i}) \eeq
is a rotation of spin $i$ about an axis
$\vec{n}_{i}=(x_{i},y_{i},z_ {i})\propto\vec{B}+\vec{B}_{{\rm
nuc},i}$ of angle $t \omega_i$ where

$\omega_{i}=\gamma_e |\vec{B}+\vec{B}_{{\rm nuc},i}|/2$.

If the system is prepared by RAP in the state
$\ket{S(t=0)}=(\ket{\upp\downn}-\ket{\downn\upp})/\sqrt{2}$, and
subsequently measured using RAP to distinguish the singlet and
triplet subspaces, the measurement probes the state's
autocorrelation function. Starting in a singlet at $t=0$, the
probability of remaining a singlet after a time $t$ is given by
the autocorrelation function
\begin{eqnarray}
A_{S}(t,B) & = & |\braket{S(t)}{S(t=0)}|^{2} \label{e:corfcn} \\
& = &
|\cos(\omega_{l}t)\cos(\omega_{r}t)+\vec{n}_{l}\cdot\vec{n}_{r}\sin
(\omega_{l}t)\sin(\omega_{r}t)|^{2}  \nonumber
\end{eqnarray}
To obtain the signal in the quasi-static approximation,
Eqn.~\ref{e:corfcn} must be averaged over the different possible
nuclear field values.  We examine the zero-field and finite-field
cases.

When $\vec{B}=0$, the properties of $\vec{n}_{i}$ within the QSA
are described by
$\mean{n_{i,\mu}n_{j,\nu}}=\delta_{ij}\delta_{\mu\nu}/3$.
Averaging over nuclei, the signal is
\ba A_{S}(t,0) & = & \mean{\cos^{2}(\omega_{l}t)}_{\rm
   nuc}\mean{\cos^{2}(\omega_{r}t)}_{\rm nuc}+ \nonumber \\
& & \ \ \frac{1}{3}\mean{\sin^{2}(\omega_{l}t)}_{\rm
nuc}\mean{\sin^ {2}(\omega_{r}t)}_{\rm nuc} \ea
where
\begin{eqnarray*}
\mean{\cos^{2}(\omega_{i}t)}_{\rm nuc} & = &
\frac{1}{2}[1+e^{-\frac
{1}{2}(t/T_{2,i}^{*})^{2}}(1-(t/T_{2,i}^{*})^{2})]~,\\
\mean{\sin^{2}(\omega_{i}t)}_{\rm nuc} & = &
\frac{1}{2}[1-e^{-\frac
{1}{2}(t/T_{2,i}^{*})^{2}}(1-(t/T_{2,i}^{*})^{2})]~.
\end{eqnarray*}
We recall that $T_{2,i}^* = (\gamma_e B_{{\rm nuc},i})^{-1}$.  A
distinct difference of this model from other dephasing mechanisms
is the order 10\% overshoot of the decay at short times, and the
asymptoptic approach of $A_{S}(t\gg T_{2}^{*},0)$ to 1/3.  A
classical master equation would exhibit neither of these
features---they are unique identifiers of the quasi-static regime,
in which different coherent dynamics are averaged over many
realizations.  Numerically we find that these qualitative features
do not depend on the relative size of the two quantum dots
($B_{{\rm nuc},l}/ B_{{\rm nuc},r}$) for variations of up to 50\%.

Another regime of interest is when the external field is much
larger than the effective nuclear fields ($|\vec{B}|\gg B_{{\rm
nuc},i}$). Spin-flip terms are highly suppressed and the system is
restricted to two levels, $\ket{S}$ and the $m_{s}=0$ triplet,
$\ket{T_{0}}=(\ket{\upp\downn}+\ket{\downn\upp})/\sqrt{2}$.  This
is described by the Hamiltonian of Eqn.~\ref{e:ms0}. This
effective two-level system's evolution operator is straightforward
to evaluate:

\beq A_{S}(t,B\gg B_{\rm nuc})=\mean{\cos^{2}(\hat{\omega}
t)}=\frac{1}{2} [1+e^{-(t/T_{2,{\rm eff}}^{*})^{2}}]\ .
\label{e:highfield} \eeq
with $T_{2,{\rm
     eff}}^{*}=1/\sqrt{\frac{1}{2}[(T_{2,l}^{*})^{-2}+(T_{2,r}^{*})^
{-2}]}$.
     Qualitatively, the decay of
the autocorrelation function $A_{S}$ due to the nuclear field is
described by Gaussian decay with a timescale $T_{2,{\rm eff}}^*$.
Similar to the case of zero magnetic field, the behavior of this
autocorrelation function is independent of variations in dot size
up to $\simeq 50\%$.

We now indicate how {\em slow} adiabatic passage at large external
magnetic field allows measurement of the results of an exchange
gate. In particular, SAP allows for preparation and measurement of
the individual spin eigenstates, $\ket{1/2,-1/2}$ and
$\ket{-1/2,1/2}$. An exchange gate leads to partial rotation
between these states, where the rotation angle is given by the
product of the exchange energy during the gate, $J(\epsilon)$, and
the time the exchange energy is non-zero, $t_E$.  Finally,
reversing SAP takes the lower energy eigenstate ($\ket{s,-s}$)
back to $\ket{(0,2)S}$ while the higher energy eigenstate
($\ket{-s,s}$) is mapped to $\ket{T_0}$, a (1,1) charge state.
The final measurement determines whether the final state is the
same as the initial state (the (0,2) result) or has changed to the
state with the two spins exchanged (the (1,1) result).  Thus,
preparing the state $\ket{1/2,-1/2}$ and measuring in the same
basis distinguishes the results of the exchange-based rotation of
the two-spin state.  For example, when the probability is 50\% for
either measurement result, a $\sqrt{\rm SWAP}$ gate has been
performed.  When the probability goes to 100\% of recovering the
higher energy eigenstate (measuring (1,1)), a complete swap of the
two spins has occurred (SWAP).

As before, we consider the probability of returning to the lower
energy eigenstate.  Now, however, this state is the $\ket{s,-s}$
state as described in the previous subsection.  After preparing
this state, we perform the resonant exchange gate of angle
$\theta_{E} = J t / \hbar$ where $t$ is the time spent waiting
with exchange energy $J$. This leads to a rotation of the state
$\ket{s,-s}$.  Its autocorrelation function is given by \ba
A_{s,-s}(t) & = & |\ketbra{s,-s(t)}{s,-s}|^2 \\
& = & \cos^2 (\theta_{E}/2)\ . \ea If the exchange term $J \leq
\hbar \gamma_e B_{\rm nuc}$, then additional effects due to nuclei
would be observed; we evaluate these below.

We emphasize that the combination of RAP  for preparation
(prepares $\ket{S}$), SAP for preparation (prepares lower energy
eigenstate of the nuclear field, $\ket{\pm}$), RAP for measurement
(spin-to-charge in $\ket{S},\ket{T_0}$ basis) and SAP for
measurement (spin-to-charge in current eigenstates of the nuclear
field, $\ket{\pm}$), when combined with the exchange gate
(rotations of $\ket{\pm}$ to $\ket{\pm} + i \ket{\mp}$) allows for
full state tomography in the $\ket{S},\ket{T_0}$ subspace.

\subsection{Errors in exchange gates}

The primary error in exchange gates is likely due to charge-based
dephasing and is directly related to the parametric dependence of
the exchange energy, $J$, on gate voltages near the charge
transition \cite{coish05,hu05,stopa06}.  In addition, other errors
are possible due to the stochastic nature of the nuclear field.
For example, there is the possibility that the current value of
the field gradient frequency, $\hat{\omega}$, is sufficiently
small to make the initial and final transfer stages non-adiabatic.
Also, the gradient can flip sign in the course of the experiment.
Finally, finite residual exchange interaction during SAP reduces
effectiveness.  We consider each of these in turn below.

In the far-detuned regime, the energy gap between the (1,1)
singlet/triplet space and higher orbital states, as well as (2,0)
and (0,2) charge states, is large (of order $\hbar \omega_0$, the
orbital level spacing of a single dot, and $E_c / 2$, the single
dot charging energy, respectively).  At dilution refrigerator
temperatures, this energy gap is many times greater than $k_b T$,
and absorption of a quanta of energy from the environment leading
to incoherent excitation may be neglected.  Also in this regime,
the residual exchange splitting  is both numerically small and
insensitive to first order fluctuations in detuning, $\epsilon$,
leading to little charge-based dephasing due to differential
coupling of the $\ket{S}$ state to the doubly-occupied states when
compared to the $\ket{T_m}$ states' couplings to doubly-occupied
states. The system remains sensitive to charge-based dephasing up
to second order due variations in the tunnel coupling, $\Tc$. If
we can write variations of $\Tc$ from the mean as $\delta \Tc$,
its correlation function is given generally by
\beq \mean{\delta \Tc(u+ \tau) \delta \Tc(u)} = \hbar^2 \int
d\omega\ S_ {\Tc}(\omega) e^{i \omega \tau}\ . \eeq

The corresponding phase noise term in the Hamiltonian  is $V_{11}
= \eta \delta \Tc(u) (\sum_m \ketbrad{T_m}-\ketbrad{S})/2$ where
$\eta = \frac{8 \Tc J}{E_c^2} \lesssim 10^{-3}$, again using $E_c
= 5$ meV, $J = 0.3$ meV, and $\mean{\Tc} = 0.01$ meV.

\begin{figure}
\centering
\includegraphics[width=\3]{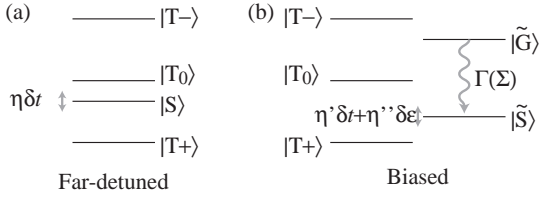}
\caption[Dephasing in the far-detuned and charge transition
regimes]{ Level structures of the double dot system in the (a)
far-detuned regime, with charge dephasing shown, and (b) near the
charge transition with charge-based decay and dephasing.
\label{f:chgdephase} }
\end{figure}

As an example, a coherence between the $\ket{S}$ and $\ket{T_m}$
subspace in the far-detuned regime could be expected to decay due
to the noise on $\Tc$ according to
\beq \mean{e^{-\frac{i \eta}{\hbar} \int_0^\tau du' \delta
\Tc(u')}} = \exp(-\eta^2 \int d\omega\ S_{\Tc}(\omega)
\frac{\sin^2(\tau \omega/ 2)}{(\omega/2)^2}) \label{e:noise} \eeq
While we consider a variety of noise sources in
Appendix~\ref{a:dephasing}, it is instructive here to take the
case of white noise spectrum with $S_{\Tc} = \gamma_0/2 \pi$. It
leads to exponential decay of coherences between $\ket{S}$ and
states of the $\ket{T_m}$ subspace with a constant $\gamma_{\Tc}
=\eta^2 \gamma_0 $. In general, as $\eta \ll 1$, this decay will
be negligible.

The charge transition  will have stronger dephasing when compared
with the far-detuned regime.  In addition to inelastic decay of
the excited adiabatic state to the ground adiabatic state, the
system is susceptible to fluctuations in both $\epsilon$ and
$\Tc$, as $J$ is potentially large and strongly dependent on gate
parameters. In so far as the power spectra associated with $\Tc$
and $\epsilon$ have no appreciable spectral components at
frequencies of order $\Epsilon$, the excitation/relaxation terms
between $\ket{\tilde{G}}$ and $\ket{\tilde{S}}$  may be neglected.
Then we may restrict considerations to dephasing of coherence
between the subspace of $\ket{T_m}$ and the state
$\ket{\tilde{S}}$, and write the effective Hamiltonian as
\beq V_{11-02} = [\eta' \delta \Tc(\tau) + \eta'' \delta
\epsilon(\tau)] [\sum_m \ketbrad{T_m}-\ketbrad{\tilde{S}}]/2 \eeq
with $\eta' = \frac{2 \Tc}{\epsilon/2+\Epsilon}$ and $\eta'' =
\frac{1}{2}(1 + \frac{\epsilon}{\epsilon/2+\Epsilon})$.  The
accompanying noise is given again by Eqn.~\ref{e:noise}, but with
$\eta=\sqrt{(\eta')^2 + (\eta'')^2}$ and the power spectrum
replaced with
\beq S_{11-02}(\omega) = \left(\frac{\eta'}{\eta}\right)^2
S_{\Tc}(\omega) +  \left(\frac{\eta''}{\eta}\right)^2
S_{\epsilon}(\omega) \eeq
While $\eta'$ may be small ($\eta \ll 1$), near the charge
transition $\eta''$ is of order unity. This indicates that noise
in the detuning parameter $\epsilon$ has significant repercussions
for coherences between singlet and triplet states during exchange
gate operation.

We consider charge-based dephasing for the exchange gate according
to Eqn.~\ref{e:noise}.  In all cases we assume the spectral
function has a high-frequency cutoff $\gamma$ such that $\gamma
\ll J$. This assumption prevents dephasing noise from producing
population changes (relaxation) due to energy conservation.
Additionally, we rewrite the expected probability of maintaining
phase coherence (Eqn.~\ref{e:noise}) as
\beq \exp(-\eta^2 P) \eeq
where $P = \int d\omega\ S(\omega)\frac{\sin^2(\tau_E
   \omega/2)}{(\omega/2)^2}$ is set by the spectral function and the
time of the exchange gate, $\tau_E$. This allows separation of the
interaction strength ($\eta$) and the noise spectrum. Each
spectrum considered has a normalization parameter $\nu$ such that
$S(\omega)$ has units of inverse time.

We note that in general, the number of observable exchange
oscillations will be limited by these dephasing processes.  By
finding $T_2 = P^{-1}(\eta^{-2})$ the observable number of
oscillations goes as $T_2 J/\hbar$.  When, for example, $S(\omega)
= \Gamma/2 \pi$, we may easily invert $P(t)$ and find $T_2 =
\eta^{-2} /\Gamma$.

\begin{figure}
\centering
   \includegraphics[width=\4]{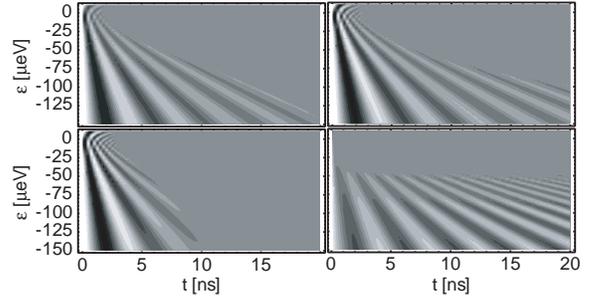}
   \caption[Decay of exchange oscillations]{Decay of exchange
oscillations for the four scenarios (clockwise from
     lower left):    white noise, $1/f$, ohmic, and super-ohmic $
(\omega^2)$.
     White corresponds to probability 1 of ending in the  initial state
     $\ket{s,-s}$ after exchange interaction is on for a time $\tau_E$
     (bottom axis),
     while black is probability 0 of ending in the initial state.
Tunnel coupling is taken to be 10 $\mu$eV, and the spectral
density ($\nu$) is chosen for similar behavior near $\epsilon =
50\ \mu$eV. Note that $1/f$ terms increase decay in the slow
oscillation limit, while increasing powers of $\omega$ (white
noise=0, ohmic = 1, super-ohmic = 2) lead to more oscillations for
smaller exchange energies. }
   \label{f:decay}
\end{figure}

Comparing the behavior of the ohmic  and super-ohmic cases to the
$1/f$ case (see Appendix~\ref{a:dephasing}), the limiting value of
$P$ for the super-ohmic case and the power law tail of the ohmic
decay indicates that for small coupling parameter $\eta$, the
super-exponential $1/f$ terms (going as a gaussian) will dominate
at long times. For very short interaction times, all three will be
less than the white noise contribution, which goes linearly in
time.  The different behaviors are shown in Fig.~\ref{f:decay}.

This indicates that electrical control of exchange interactions in
double dot systems may be relatively robust with respect to
nuclear spin degrees of freedom.  However, during the exchange
gate, the system is susceptible to charge-related dephasing. The
observed decay of oscillations of Ref.~\onlinecite{petta05}, in
which the decay rate appears to match the exchange energy such
that the number of exchange oscillations observed is independent
of the detuning, is qualitatively similar to the behavior of
oscillations in the presence of sub-ohmic noise.  A more detailed
experimental analysis of the noise will be required before a
direct comparison between experiment and theory will be possible.


\section{Exchange gates and echo techniques \label{s:echo}}

The techniques of rapid adiabatic passage, slow adiabatic passage,
and time-resolved control of the exchange interaction in the
previous section have far reaching applications for quantum
control of spins \cite{hanson07}. In this section, we consider how
exchange gates can undo the effect of the quasi-static nuclear
field, greatly reducing the deleterious effects of nuclear spins
on electron spin coherences and allowing for electron spin
measurements to determine nuclear spin correlation functions.

\subsection{Spin-echo in the singlet-triplet basis}

Since the nuclear fields vary slowly on timescales compared to a
typical pulse cycle time, a spin-echo pulse sequence can be used
to refocus the spin singlet state. A spin-echo pulse sequence
based on fast electrical control of the exchange interaction was
demonstrated in Ref.~\onlinecite{petta05}. The experiment starts
by using RAP to transfer $\ket{(0,2)S}$ to $\ket{S}$, preparing a
separated singlet pair of electron spins.  As demonstrated in the
$T_2^*$ experiment, the hyperfine interaction mixes the singlet
and triplet states on a 10 ns time scale. After a separation time
$t_S$, a $\pi$ exchange pulse (SWAP) is performed by adjusting the
detuning to a region with a finite exchange energy. The exchange
energy is then set to zero by moving to the far-detuned regime for
a time $t_{S'}$, during which the singlet state refocuses.

The dephasing due to hyperfine interactions occurs by producing a
relative, unknown phase between $\ket{\upp \downn}$ and
$\ket{\downn \upp}$. Switching between these two states via
exchange gates will produce an echo (recovery of the original
state, $\ket{S}$) if the waiting time before and after the $\rm
SWAP$ operation is the same.  In other words, the pseudo-spin of
$\ket{S}$ and $\ket{T_0}$ (the $m_s=0$ subspace) is amenable to
echo techniques using exchange interactions.

We will use the pulse sequence of Fig.~\ref{f:pulses}c.  In the
far-detuned regime at finite field, $\ket{S}$ and $\ket{T_0}$ mix
due to nuclei, as per the Hamiltonian of Eqn.~\ref{e:ms0}.  This
mixing is driven by a relative constant, but unknown, energy,
corresponding to the current value of
\beq \hbar\hat{\omega}=\gamma_e (\hat{B}_{{\rm
nuc},l,z}-\hat{B}_{{\rm nuc},r,z})~. \eeq
Within the corrections to the quasi-static approximation, $\hat
{\omega} = \gamma_e (\delta \hat{B}_{l,z}(t) - \delta
\hat{B}_{r,z}(t) $, where $\delta \hat{B}_{l(r),z}(t)$ are random,
Gaussian variables described by Eqn.~\ref{e:powerspec}.  For
clarity, we rewrite it here:
\begin{equation}
\mean{\hat{\omega}(t+\tau) \hat{\omega}(t)} = \int d\omega\
S(\omega) e^{i \omega \tau}
\end{equation}

For example, the singlet-singlet correlation function at finite
field (Eqn.~\ref{e:highfield}) is modified, noting that
\begin{equation}
\mean{\cos^2(\int_0^t \hat{\omega}(t') dt')} = \frac{1}{2}\left[1
+ \exp(-\int d\omega\ S(\omega) \frac{\sin^2(\tau
\omega/2)}{(\omega/2)^2}) \right]
\end{equation}
For $S(\omega)$ with a high frequency cutoff below $1/\tau$, we
can Taylor expand the $\sin$ term.  Then, comparison with
Eqn.~\ref {e:highfield} indicates:
\begin{equation}
T_2^* = \sqrt{\int d\omega\ S(\omega)}
\end{equation}

Now, we consider how this result changes with the more complex
sequence given by Fig.~\ref{f:pulses}c, in particular its
dependence on $\tau_{E}$ and $\tau_{S'}$. If $\tau_{E}=0$, then
nothing has changed from before, except now the system evolves
according to $U_{\rm nuc}(\tau_{S'})U_{\rm nuc}(\tau_{S})=U_{\rm
nuc}(\tau_{S}+\tau_ {S'})$. However, what happens with finite
$\tau_{E}$?

For our reduced two-level system, when $\tau_{E}\neq0$ the
effective Hamiltonian during this stage is given by \beq
H_{eff,E}=\hbar\hat{\omega}\hat{\sigma}_{x}/2+J(\epsilon_E)\hat
{\sigma}_{z}/2\ . \eeq where Pauli matrices with states
$\ket{S},\ket{T_{0}}$ as a psuedo- spin, in our logical basis
defined above (i.e., $\sigma_z = \ketbrad{T_0} - \ketbrad{S}$).
Taking the turn-on of finite $J$ to be instantaneous, the total
evolution operator is \beq U(\tau_{S'},\tau_{E},\tau_{S})=U_{\rm
nuc}(\tau_{S'})\exp(-iH_{eff,E}\tau_{E}/\hbar)U_{\rm
nuc}(\tau_{S}) \ . \eeq When the exchange energy for the middle
point satisfies $J(\epsilon_E) \gg \gamma_e B_{\rm nuc}$, we
approximate the middle term of the evolution operator by
$U_{E}(\tau_{E})\approx\exp(-i\tau_{E}J(\epsilon_{E})\hat{\sigma}_{z}/
2\hbar)$.

For now, taking $\hat{\omega}$ to be constant, at the end of the
pulse sequence, the probability of returning to the state
$\ket{(0,2)S}$ ($C$) is given by
\beq C = {\rm Tr}_{{\rm
nuc}}\left[\left|\bra{0}e^{-i\tau_{S'}\hat{\omega}
\hat{\sigma}_{x}/2}e^{-i\tau_{E}J(\epsilon_{E})\hat{\sigma}_{z}/2
\hbar}e^{-i\tau_{S}\hat{\omega}\hat{\sigma}_{x}/2}\ket{0}\right|^{2}
\right]~. \eeq
To see the analogy between this evolution and a spin echo
experiment, we insert unity after the initial ket, i.e.,
$\ket{0}=e^{iJ(\epsilon_{E})\hat{\sigma}_{z}/2\hbar}e^{-iJ
(\epsilon_{E})(-1)/2\hbar}\ket{0}$, as 0 is the -1 eigenstate of
$\sigma_{z}$. The overall phase is irrelevant due to the absolute
value terms, and so we rewrite the above as
\begin{widetext}
\beq C={\rm Tr}_{{\rm
nuc}}\left[\left|\bra{0}e^{-i\tau_{S'}\hat{\omega}
\hat{\sigma}_{x}/2}\left(e^{-i\tau_{E}J(\epsilon_{E})\hat{\sigma}_{z}/
2\hbar}e^{-i\tau_{S}\hat{\omega}\hat{\sigma}_{x}/2}e^{i\tau_{E}J
(\epsilon_{E})\hat{\sigma}_{z}/2\hbar}\right)\ket{0}\right|^{2}\right]~.
\eeq
\end{widetext}
The term in parenthesis, an exchange gate, is a rotation of angle
$\theta_{E}=\tau_{E}J(\epsilon_{E})/\hbar$ about the $z$ axis of
the psuedo-spin.  It acts on the operator
$U_{nuc}(\tau_{S})=e^{-i\tau_{S}\hat{\omega}\hat{\sigma}_{x}/2}$,
flipping the sign of the $\sigma_x$ operator when $\theta_{E} =
\pi$ (SWAP).

We can probe its effects by analogy with standard spin echo. For
example, when $\theta_{E}=\pi$, we get
$C=\frac{1}{2}[1+\exp(-\mean{\omega^{2}}(\tau_{S}-\tau_{S'})^{2}/2)]$.
This means the dephasing due to nuclei in the first half of the
sequence is exactly undone when the rephasing time is equal
($\tau_S = \tau_{S'}$). For fixed $\tau_{S}+\tau_{S'}$, the
probability of returning to the singlet state as a function of
$\tau_{S}-\tau_{S'}$ and $\tau_ {E}$ should exhibit 50\% mixing
expected for dephasing when $|\tau_{S}-\tau_{S'}|  > T_2^*$, and
when $\tau_{E} J/\hbar$ is not an odd integer multiple of $\pi$.
For example, setting $\tau_{S}+\tau_{S'}=100$ ns, and using
$\epsilon_{E}$ such that $J(\epsilon_{E})=\hbar \gamma_e (20\ {\rm
mT})$, the probability exhibits this behavior.

In practice, the instantaneous approximation breaks down in
realistic situations, as does the $J \gg \gamma_e B_{\rm nuc}$
assumption.  The former is easily fixed by noting that such a
``Rabi'' type pulse only has sensitivity to the integrated area,
i.e., $\theta = \int J(\epsilon(t)) dt$. The latter requires
working with finite $\tau_{E}\Omega$.  This effect has been
considered in detail elsewhere~\cite{taylor05spinbath}.

\subsection{Probing nuclear spin dynamics through echo techniques
\label{ss:echo}}

So far the analysis has worked entirely within the QSA.  However,
the echo sequence in principle reveals higher order dynamical
information about the nuclear field and other noise sources.
Effects of nuclear spin dynamics on electron spin decoherence has
been considered by several authors
\cite{desousa03,coish04,yao05,witzel05,shenvi05}. We now consider
the echo sequence with $\hat{\omega}(t)$ a Gaussian variable
slowly varying in time. This allows a large range of possible
noise sources and correlation functions to be considered.

Assuming (for the moment) that the exchange gate is of high
fidelity and insensitive to the current value of $\Az$, the
measured result of the echo sequence is given by
\begin{eqnarray}
C & = & |\bra{0_{L}}\exp(i\int_{t_{S}}^{t_{S}+t_{S'}}
\hat{\omega}(t)
dt\sigma_{x}/2 \nonumber \\
& & \ \
-i\int_{0}^{t_{S}}\hat{\omega}(t)dt\sigma_{x}/2)\ket{0_{L}}|^
{2}\\
& = &
\cos^{2}[(\int_{t_{S}}^{t_{S}+t_{S'}}\hat{\omega}(t)dt+\int_{0}^
{t_{S}} \hat{\omega}(t) dt)/2]\\
& = & \frac{1}{4}[e^{-i \Xi}+e^{i \Xi}+2]\\
& = & \frac{1}{2}[e^{-\mean{\Xi^{2}}/2}+1]
\end{eqnarray}
where $\Xi=\int_{t_{S}}^{t_{S}+t_{S'}}
\hat{\omega}(t)dt+\int_{0}^{t_ {S}} \hat{\omega}(t)dt$. The second
moment of $\Xi$ is \beq
\mean{\Xi^{2}}=\mean{\Xi_{S'}^{2}}+\mean{\Xi_{S}^{2}}-2\mean{\Xi_{S'}
\Xi_{S}} \eeq with $\mean{\Xi_{S'}^{2}}=\int d\nu\
S(\nu)\frac{4}{\nu^{2}}\sin^{2}
(t_{S'}\nu/2)\simeq(t_{S'}/T_{2}^{*})^{2}$ and similarly for
$\Xi_{S} $. The cross term, corresponding to the correlations
between the two frequencies, is \beq 2\mean{\Xi_{S'}\Xi_{S}}=\int
d\nu\  S(\nu)\frac{\sin(t_{S}\nu/2)\sin
(t_{S'}\nu/2)}{(\nu/2)^{2}}e^{i(t_{S}+t_{S'})\nu/2}~. \eeq
Finally, when $t_{S}=t_{S'}$, the second moment is \beq
\mean{\Xi^{2}}=2\int d\nu\
S(\nu)\frac{\sin^{2}(t_{S}\nu/2)}{(\nu/2)^{2}}[1-e^{it_{S}\nu}]~.
\eeq For low frequency noise, with cutoff $\gamma\simeq[\int d\nu\
\nu^{2}S(\nu)]^{1/2}\ll1/t_{S}$, we obtain
$\mean{\Xi^{2}}\simeq2(t_{S})^{2}\times(t_{S}\gamma)^{2}$, or
decay in the total wait time $2t_{S}$ with an effective time
constant \beq T_{2,SE}=8^{1/4}(T_{2}^{*}/\gamma)^{1/2}\ . \eeq Our
evaluation has implicitly assumed that above the cutoff, $\gamma$,
$S(\omega)$ dies off at least as $1/\omega^3$.  We suggest that
this is appropriate for interaction times on the order of
microseconds---initial decay of the nuclear spin correlation
function is quadratic.  For longer interaction times, a different
decay morphology (going as $\exp(-t^3)$) could be observed.

Our predicted decay is consistent with the experiments of
Ref.~\onlinecite{petta05}.  In
Refs.~\onlinecite{witzel05,yao05,deng05,shenvi05}, $\gamma$ is of
order 10 ms$^{-1}$, giving $T_{2,SE} = 2\ \mu$s.  However,
addition exponential decay could be observed if the high frequency
cut-off assumed above has a too-slow decay, going as
$\omega^{-2-\epsilon}$ with $\epsilon < 1$.  Furthermore, higher
order pulse sequences, such as Carr-Purcell, will likely allow for
extensions of the echo signal to substantially longer
times\cite{witzel07,zhang07,bhakta07}.

The experiments discussed in the previous sections demonstrate
that the hyperfine interaction is efficient at dephasing an
initially prepared spin singlet state on a 10 ns timescale. By
using a simple spin-echo pulse sequence, this bare dephasing time
was extended by over a factor of 100, to times $T_2$$>$1.2$\mu$s.
Further experimental effort will be required to fully map out the
nuclear correlation function and extend the lower bound on
electron spin coherence times.

\section{Conclusions}

We have shown how a model combining charge and spin interactions
for two-electrons in a double quantum dot effectively describes
the experimental results of Refs.~\onlinecite{johnson05,petta05}.
By starting with the case of a single electron in a single dot, we
employed the quasi-static approximation
\cite{burkard99,khaetskii02,coish04,shenvi05,coish05,witzel05,yao05,deng05},
and considered its applicability to describing current
experimental results.

The spin interactions with nuclear spins are
extended to the double dot case, and two regimes emerge: the far-detuned
regime, in which two electrons are in separate dots and interact with
independent nuclear fields, and the charge transition, in which the two
electrons may transition from a separated orbital state to a
doubly-occupied, single-dot state.

This model was used to describe spin blockade, and we found spin
blockade is broken by interactions with nuclei near zero magnetic
field, explaining the experimental results of Johnson {\em et al.}
(Ref.~\onlinecite{johnson05}).  A striking magnetic field
dependance is derived, consistent with observed experimental
behavior.  This indicates that the dominant mechanism for spin
blockade at finite magnetic field is trapping of the $m_s=\pm 1$
separated spin states, as their mixing with the
charge-transition-allowed singlet $\ket{G}$ is suppressed by
finite Zeeman splitting.  Observation of the breaking of spin
blockade near zero field provides a sensitive measure of the
magnitude of the random nuclear-spin-induced magnetic field.

Time-domain control of local potentials, achieved through
manipulation of  electrostatic depletion gates, provides powerful
mechanisms for preparing and measuring spin singlets, as well as
eigenstates of the nuclear field.  These techniques have been
exploited by Petta \textit{et al.} to measure the effective
dephasing of a two-spin entangled  state, and to probe via
coherent oscillations in the $m_s=0$ two-spin subspace the
exchange interaction as controlled by gate voltages
\cite{petta05}. Limiting mechanisms for such oscillations, due to
charge fluctuations of indeterminate nature, are considered for a
variety of environmental noise spectra.

Finally, we analyzed how controlled exchange interactions can
protect the electron spin from the deleterious effect of nuclear
spins by working within a two-spin subspace, putting in specific
terms protocols previously conceived more generally.  We expect
the limiting mechanism for the rephasing of the two-spin states
comes from corrections to the quasi-static approximation, and as
such, spin-echo experiments provide a useful measure of the
validity of this approximation.

\begin{acknowledgments}
  The authors would like to thank H.-A.\ Engel, W.\ D\"ur, P.\ Zoller,
  G.\ Burkard, and D.\ Loss for helpful discussions.  This work was
  supported by ARO, NSF, Alfred P. Sloan Foundation, David and Lucile
  Packard Foundation.
\end{acknowledgments}

\appendix

\section{Adiabatic elimination for nuclear-spin-mediated inelastic decay \label{s:inelastic}}

We will transform the superoperator (Eqn.~\ref{e:L}) into the interaction picture, but first introduce matrix elements between the eigenstates of the quasi-static fields and the state $\ket{S}$ occurring in the superoperator.  For a single spin in a magnetic field $\vec{B} = B_{\rm nuc} (x,y,z)$ and $|B| =  B_{\rm nuc} n$ (the roman variables $x,y,z,n$ are chosen such that the nuclear field contribution will be of order unity), the eigenstates may be written by rotation from spin states aligned with the $z$-axis ($\{\ket{\upp}, \ket{\downn} \}$):
\begin{widetext}
\beq
\left( \begin{array}{c}
 \ket{1/2} \\
 \ket{-1/2}
\end{array} \right)
=
\lim_{x'\rightarrow x^+}
\frac{1}{\sqrt{2n (n+z)}}
\left(
\begin{array}{cc}
n+z & x' - i y \\
x' + i y & -n-z
\end{array} \right)
\left(
\begin{array}{c}
\ket{\upp} \\
\ket{\downn}
\end{array} \right)
\eeq
The limit is taken only to remove the degenerate case of field anti-aligned with the $z$-axis, i.e., $\vec{B}=(0,0,-B)$, which would be degenerate for this matrix, and is implicit in what follows. The corresponding eigenvalues of the Hamiltonian are $\pm \hbar \Omega n/2$ in this notation.

Setting $\ket{S} = (\ket{\upp \downn} - \ket{\downn \upp})/\sqrt{2}$ and using the single spin transformations for $l$ and $r$ separately (with $\vec{B}^l = B_{\rm nuc}^l (x_l,y_l,z_l), |B^l| = n_l$ and similarly for $r$), we write
\ba
c_{1/2,1/2} = \braket{\frac{1}{2},\frac{1}{2}}{S}  & = &
    \frac{1}{\mathcal N}\left[ (x_l + i y_l)(n_r + z_r) - (n_l + z_l)(x_l+i y_l) \right] \\
c_{1/2,-1/2} = \braket{\frac{1}{2},-\frac{1}{2}}{S}  & = &\frac{1}{\mathcal N}
    \left[-(n_l+z_l)(n_r+z_r) - (x_l + i y_l)(x_r - i y_r) \right] \\
c_{-1/2,1/2} = \braket{-\frac{1}{2},\frac{1}{2}}{S} & = & -c_{1/2,-1/2}^* \\
c_{-1/2,-1/2} = \braket{-\frac{1}{2},-\frac{1}{2}}{S} & = & c_{1/2,1/2}^* \\
{\rm and}\ {\mathcal N} & = & \sqrt{8n_l  n_r (n_l+z_l)(n_r+z_r)}
\ea
It is convenient to define $c_{++} = c_{1/2,1/2}$ and $c_{+-} = c_{1/2,-1/2}$ as the spin-aligned and spin-anti-aligned coefficients, respectively.  This allows us to express $\ket{S}$ occurring in Eqn.~\ref{e:L} in terms of the eigenstates of the Hamiltonian as
\beq
\ket{S} =  \sum_{s,s'} c_{s,s'} \ket{s,s'}  = c_{++} \ket{\frac{1}{2},\frac{1}{2}} + c_{+-} \ket{\frac{1}{2},-\frac{1}{2}} - c_{+-}^* \ket{-\frac{1}{2},\frac{1}{2}} + c_{++}^* \ket{-\frac{1}{2},-\frac{1}{2}}
\eeq
In the interaction picture, each eigenstate $\ket{s,s'}$ evolves according to
\beq
\ket{s,s'(t)} = e^{-i E_{s,s'} t/\hbar}\ket{s,s'} = e^{-i (s n_l
\gamma_e B_{{\rm nuc},l} + s' n_r \gamma_e B_{{\rm nuc},r}) t}\ ,
\eeq and $\ket{S(t)} = \sum_{s,s'} c_{s,s'} \ket{s,s'(t)}$.

With these results, the louivillian may be put into the interaction picture:
\ba
\dot{\tilde{\rho}} & = & \Gamma(\Epsilon)/2  \sum_{s,s',r,r'} c_{s,s'} c_{r,r'}^* e^{i [(r-s) n_l \gamma_e B_{{\rm nuc},l} + (r'-s') n_r \gamma_e B_{{\rm nuc},r}] t} L_{s,s',r,r'}[\tilde{\rho}] \label{e:Lint}\\
{\rm where}\ L_{s,s',r,r'}[\tilde{\rho}] & = & \left[
\ketbra{s,s'}{r,r'} \tilde{\rho} +\tilde{\rho}  \ketbra{s,s'}{r,r'}  - 2 \ketbra{G}{r,r'} \tilde{\rho} \ketbra{s,s'}{G} \right]
\ea
So far, this result is exact within the QSA.

When $\Gamma(\Epsilon) \ll \gamma_e B_{{\rm nuc},l}, \gamma_e B_{{\rm nuc},r}$, the exponential phase terms of Eqn.~\ref{e:Lint} oscillate substantially faster than $\tilde{\rho}$ evolves. Adiabatic elimination becomes an appropriate approximation when we may neglect quickly rotating terms, i.e., if we may neglect certain degenerate cases, such as situations with $|n_l \gamma_e B_{{\rm nuc},l} - n_r \gamma_e B_{{\rm nuc},r}| \lesssim \Gamma(\Epsilon)$.  In addition, we are implicitly assuming that $\Gamma(\Epsilon) \simeq \Gamma(\Epsilon + E_{s,s'}-E_{r,r'})$, which is appropriate for large $\Epsilon$ and smooth phonon density of states.

More explicitly, we can average each term of over several spin rotations and make a Born approximation:
\beq
e^{i [(r-s) n_l \gamma_e B_{{\rm nuc},l} + (r'-s') n_r \gamma_e B_{{\rm nuc},r}] t} L_{s,s',r,r'}[\tilde{\rho}(t)] \rightarrow
\left[\frac{1}{\tau} \int_{t-\tau}^{t} e^{i [(r-s) n_l \gamma_e B_{{\rm nuc},l} + (r'-s') n_r \gamma_e B_{{\rm nuc},r}] t'}  dt' \right]  L_{s,s',r,r'}[\tilde{\rho}(t)]
\eeq
The time averaging for a given $s,s',r,r'$ is straightforward so long as $n_l B_{{\rm nuc},l} \neq n_r B_{{\rm nuc},r}$\footnote{
When integrating over nuclear spin degrees of freedom, this corresponds to a surface of measure 0.
}, giving
\beq
\lim_{\tau\rightarrow \infty} \frac{1}{\tau} \int_{t-\tau}^{t} e^{i [(r-s) n_l \gamma_e B_{{\rm nuc},l} + (r'-s') n_r \gamma_e B_{{\rm nuc},r}] t'}  dt'  = \delta_{s,r} \delta_{s',r'}
\eeq
Thus terms with quickly varying phase go to zero.

\end{widetext}

\section{Dephasing power spectra \label{a:dephasing}}

We now evaluate dephasing in exchange gates due to charge
fluctuations for a variety of spectral functions.  The error
should go as $[1 - \exp(-\eta^2 P)]/2$, where the value $\eta$
depends only on the detuning parameter.  The impact of the
particular spectral function is encompassed in $P = \int d\omega\
S(\omega)\frac{\sin^2(t
  \omega/2)}{(\omega/2)^2}$.  We have assumed that $S(\omega)$ has a
high frequency cutoff $\gamma \ll 1/t$.

\subsection{White noise}

We set $S(\omega) = \frac{\nu}{2 \pi}e^{-\omega/\gamma}$.  Then we can evaluate
\beq
P = \frac{\nu}{2 \pi} \int d\omega\ \frac{\sin^2(\omega t/2)}{(\omega/2)^2}e^{-\omega/\gamma} \approx \nu t\ .
\eeq
This indicates the expected exponential decay of coherence due to white noise dephasing.

\subsection{$1/f$ noise}

With $S(\omega) = \nu^2/\omega$ and frequency cutoffs $B < \omega < \gamma \ll 1/t$,
\beq
P = 2 \log[\gamma/B] (\nu t)^2 \ .
\eeq
For a bath of $1/f$ distributed fluctuators, the initial dephasing is
quadratic in the time of interaction, and increases as the measurement
timescale ($1/B$) increases. At long times, the decay is gaussian,
with super-exponential suppression of coherence.

\subsection{Ohmic noise}

Taking $S(\omega) = g \omega e^{-\omega/\gamma}$, evaluation of $P$ is possible, giving
\beq
P = 2 g \log[1+(\gamma t)^2]\ .
\eeq
When considered in the decay function $\exp(-\eta^2 P)$, this gives a non-exponential decay law, $\exp(-\eta^2 P) = [1+(\gamma t)^2]^{-2 g \eta^2}$.  In the short time limit, this is quadratic decay, going as $1-2(g \eta^2)(\gamma t)^2 + O((\gamma t)^4)$, while the long time behavior is a power law with power $-4 g \eta^2$.

\subsection{Super-ohmic noise}

For the final spectral function considered here, we set $S(\omega) = \nu^{1-\zeta} \omega^{\zeta} e^{-\omega/\gamma}$ where $\zeta > 1$ indicates super-ohmic noise.  Evaluation of $P$ proceeds in a straightforward manner, giving
\ba
P & = & \frac{1}{4} \Gamma(\zeta-1) \left(\frac{\nu}{\gamma}\right)^{1-\zeta}
\Big\{ 1 - \nonumber \\
& & \ \  \left[1 + (\gamma t)^{\frac{1}{2} - \frac{\zeta}{2}}\right] \cos[(\zeta-1)\tan^{-1}(\gamma t)]\Big\}\ .
\ea
where $\Gamma(x)$ is the gamma function.
This type of decay has a limiting value of
\beq
\lim_{t \rightarrow \infty} P = 4 \Gamma(\zeta-1) \left(\frac{\nu}{\gamma}\right)^{1-\zeta}
\eeq
and short time behavior according to
\beq
P(t) = 2 \zeta (\zeta-1) \Gamma(\zeta-1) \left(\frac{\nu}{\gamma}\right)^{1-\zeta} (\gamma t )^2 + O((\gamma t)^4)\ .
\eeq

For visual comparison, we calculate the expected, observable Rabi
oscillations using SAP as a function of time at finite exchange
($t_E$) and at detuning $\epsilon$ in Fig.~\ref{f:decay}.
In essence, increasing the exponent of
the noise spectra (from $1/\omega$ to constant to $\omega^\zeta$)
leads to more oscillations as detuning is made more negative, i.e., as
the admixture of charge decreases.

\end{document}